\documentclass[10p,twocolumn,pra,amssymb,nobibnotes,amsmath,aps,showpacs]{revtex4-1}

\usepackage[breaklinks=false]{hyperref}
\usepackage{bm}
\usepackage{graphics}
\usepackage{graphicx}
\usepackage{braket}
\usepackage{float}
\usepackage{multirow}
\usepackage[usenames,dvipsnames]{color}

\begin{document}%
\title[Production of optically trapped $^{87}$RbCs Feshbach molecules]
{Production of optically trapped $^{87}$RbCs Feshbach molecules}

\author{Michael P. K\"oppinger, Daniel J. McCarron, Daniel L. Jenkin,
Peter K. Molony, Hung-Wen Cho and Simon L. Cornish,}
\affiliation{Joint Quantum Centre (JQC) Durham/Newcastle, Department of
Physics, Durham University, South Road, Durham DH1 3LE, United Kingdom}

\author{C. Ruth Le Sueur, Caroline L. Blackley, and Jeremy M. Hutson }
\affiliation{Joint Quantum Centre (JQC) Durham/Newcastle, Department of Chemistry,
Durham University, South Road, Durham, DH1 3LE, United Kingdom}

\begin{abstract}
We report the production of $^{87}$RbCs Feshbach molecules in a crossed-beam
dipole trap. A mixture of $^{87}$Rb and $^{133}$Cs is cooled close to quantum
degeneracy before an interspecies Feshbach resonance at 197~G is used to
associate up to $\sim5000$ molecules with a temperature of $\sim300$\,nK. The
molecules are confined in the dipole trap with a lifetime of 0.21(1)\,s, long
enough for future experiments exploring optical transfer to the absolute ground
state. We have measured the magnetic moment of the Feshbach molecules in a
magnetic bias field range between 181 and 185\,G to demonstrate the ability to
control the character of the molecular state. In addition we have performed
Feshbach spectroscopy in a field range from 0 to 1200\,G and located three
previously unobserved resonances at high magnetic fields.

\end{abstract}

\maketitle

\section{Introduction}\label{sec:Introduction}

The production of ultracold polar molecules is of significant interest for a
wide range of potential applications \cite{Carr:NJPintro:2009}. The permanent
electric dipole moments of these molecules give rise to anisotropic, long-range
dipole-dipole interactions which, when coupled with the precise control
attainable for quantum systems, result in new prospects for research. Samples
of ultracold molecules could prove useful for quantum simulation of condensed
matter systems \cite{Micheli:2006,Barnett:2006}, quantum computation
\cite{DeMille:2002,Andre:2006}, precision measurements
\cite{Flambaum:2007,Isaev:2010,Hudson:2011,Leanhardt:2011,Baron2013} and controlled
chemistry \cite{Hudson:2006a,Krems:2008}.

Although other methods are making progress
\cite{Bethlem:IRPC:2003,Sage:2005,Aikawa:2010,Barry:2012}, currently the most
successful technique for creating ultracold polar molecules relies on a
two-step indirect method where the constituent atoms in a mixed-species quantum
gas are associated into ground-state molecules \cite{Damski:2003}. First weakly
bound molecules are made by magnetoassociation using an interspecies Feshbach
resonance. Here an applied magnetic field is swept across the resonance, such
that the energy of the separated atomic states is tuned adiabatically through
an avoided crossing with the energy of a weakly bound molecular state
\cite{Chin:RMP:2010}. The molecules are then optically transferred into their
ro-vibrational ground state by stimulated Raman adiabatic passage (STIRAP)
\cite{Bergmann:1998}. Although this method has been successfully applied in
several systems \cite{Ni:KRb:2008,Lang:ground:2008,Danzl:ground:2010},
fermionic KRb is the only polar molecule that has so far been produced at high
phase-space density \cite{Ni:KRb:2008}. However KRb is unstable as the exchange
reaction \mbox{2KRb $\rightarrow$ K$_2$ + Rb$_2$} is exothermic
\cite{Ospelkaus:react:2010}. An appealing alternative is to produce
ground-state RbCs which is expected to be collisionally stable because both the
exchange reaction and trimer formation reactions are endothermic
\cite{Zuchowski:trimers:2010}.

Pilch {\em et al.}\ \cite{Pilch:2009} have investigated Feshbach resonances in
$^{87}$RbCs at magnetic fields up to 300\,G. Subsequently, Lercher {\em et
al.}\ \cite{Lercher:2011} achieved simultaneous $^{87}$Rb and Cs BECs in
separated dipole traps, while McCarron {\em et al.}\ \cite{McCarron:2011}
achieved a dual-species BEC in a single crossed dipole trap, where a phase
separation of the condensates was observed \cite{Pattinson2013}. Most recently,
Takekoshi {\em et al.}\ \cite{Takekoshi:RbCs:2012} extended the Feshbach
spectroscopy to 667\,G and reported the production of $\sim3000$ Feshbach
molecules. They also combined information from the spectroscopy of the
$X^1\Sigma^+$ and $a^3\Sigma^-$ molecular states with the ultracold results to
obtain a precise coupled-channels model of the interaction. The interaction
potentials derived for $^{87}$RbCs subsequently gave accurate predictions of
resonance positions in $^{85}$RbCs by mass scaling \cite{Cho2013}.

In this paper we report the production of up to $\sim5000$ $^{87}$RbCs Feshbach
molecules from a nearly quantum-degenerate sample of $^{87}$Rb and $^{133}$Cs.
Molecules are trapped in  an optical dipole trap with a lifetime of 0.21(1)\,s.
We report measurements of the magnetic moment in a field range from 181 to
185\,G to demonstrate the ability to control the character of the molecular
state and to confirm the theoretical bound-state model. In the course of this
work, we have also measured the interspecies Feshbach spectrum in a magnetic
field range up to 1200\,G, focussing on the previously unexplored magnetic
field region above 600\,G.

The structure of the paper is as follows. In Sec.\,\ref{sec:Resonances} we
present the results of our work on the interspecies scattering between
$^{87}$Rb and $^{133}$Cs. Calculations of the scattering length and the binding
energy for weakly bound states are presented, together with our experimental
observations of the Feshbach spectrum. In Sec.\,\ref{sec:Molecules} we focus on
the production of $^{87}$RbCs Feshbach molecules and characterise the
near-threshold bound-state spectrum through measurements of the magnetic moment
of the molecules. In Sec.\,\ref{sec:Conclusion} we conclude with an outlook on
further experiments.

\section{The $^{87}$R\lowercase{b}C\lowercase{s} Feshbach spectrum}\label{sec:Resonances}

\subsection{Overview}

\begin{figure*}[!]
\includegraphics[width=2\columnwidth]{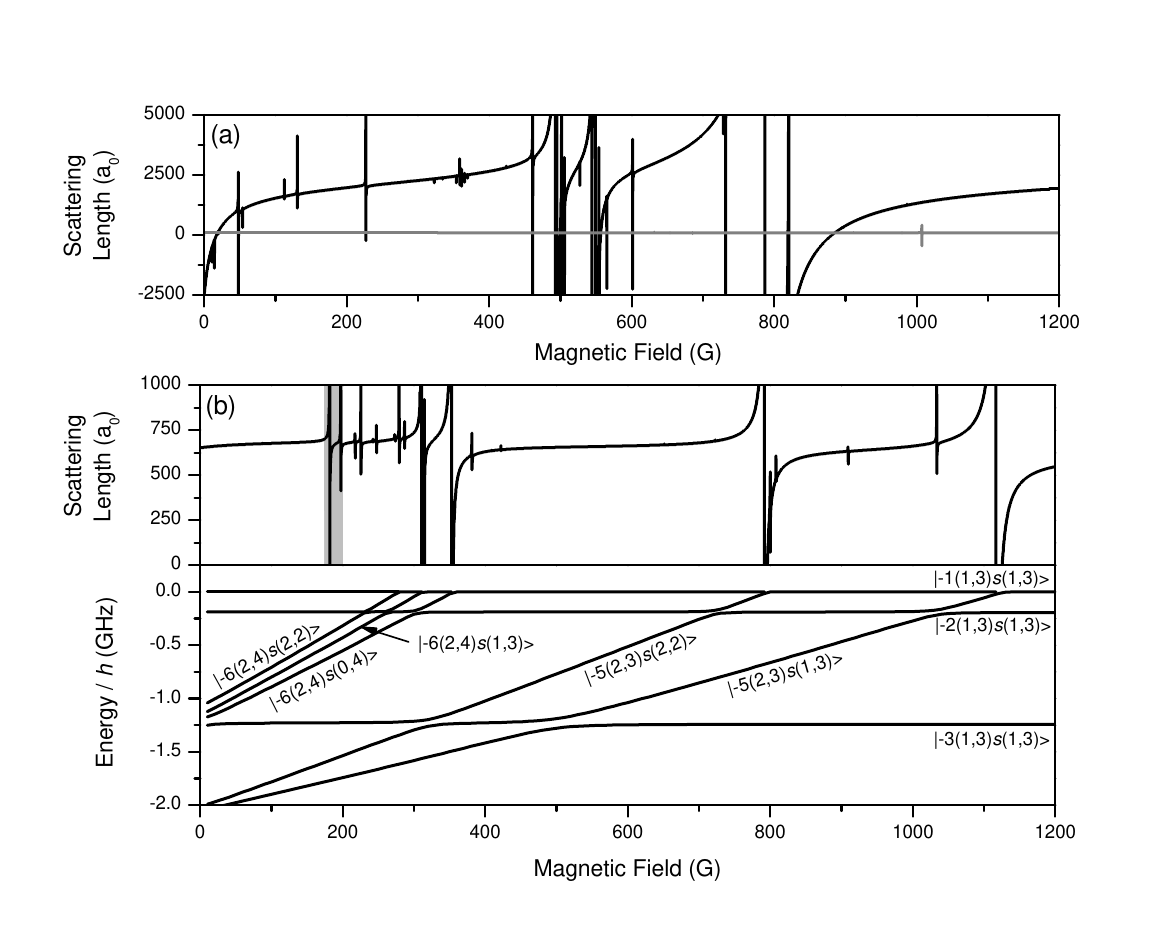}

\caption{(a) The intraspecies scattering length of $^{133}$Cs in the
$(f,m_f)=(3,+3)$ state (black) and $^{87}$Rb in the $(1,+1)$ state (grey),
calculated using the \emph{s}, \emph{d} and \emph{g} basis functions for
$^{133}$Cs and \emph{s} and \emph{d} basis functions for $^{87}$Rb versus
magnetic field. (b) Upper panel: The interspecies scattering length between
$^{87}$Rb and $^{133}$Cs in the same spin states as in (a), calculated using
the \emph{s} and \emph{d} basis functions, versus magnetic field. The grey
shaded area marks the range relevant to the magnetoassociation discussed in
Sec.~\ref{sec:Molecules} and shown in Fig.\,\ref{fig:BoundState}. Lower panel:
The calculated weakly bound molecular states arising from $L=0$
(\emph{s}-states). The bound-state energies are plotted relative to the energy
of the lowest hyperfine state of $^{87}$Rb +$^{133}$Cs, that is, the
$(1,+1)+(3,+3)$ hyperfine level. All the bound states shown are for $M_{\rm
tot}= 4$, corresponding to $s$-wave scattering in the lowest channel.}
\label{fig:sBoundspectrum}%
\end{figure*}

The properties of a quantum-degenerate mixture are strongly influenced by both
the intraspecies and the interspecies interactions
\cite{Matthews:1999,McCarron:2011,Papp2008,Becker2008}. As such, full knowledge
of all the scattering lengths is essential when devising a strategy to produce
atomic mixtures with the high phase-space densities needed for
magnetoassociation. In general, $^{87}$Rb is a desirable constituent of a
mixed-species cold atom experiment since its scattering properties make cooling
to quantum degeneracy possible over a wide magnetic field range and in several
different internal states. The scattering properties of $^{133}$Cs are very
different, exhibiting a large background scattering length and a rich spectrum
of Feshbach resonances
\cite{Leo:2000,Chin2000,Chin:cs2-fesh:2004,Berninger2013}. Efficient
evaporation to Bose-Einstein condensation has been possible only in the
absolute ground state $(f=3, m_f=+3)$ where inelastic two-body collisions
cannot occur \cite{Weber:CsBEC:2003,Hung2008,McCarron:2011}. Even in this
state, efficient evaporation is possible only for moderate scattering lengths
in the vicinity of the zero-crossing of broad Feshbach resonances at
17.12, 556.26 and 880.90\,G \cite{Berninger2013}. The intraspecies scattering
lengths for $^{87}$Rb and $^{133}$Cs in their lowest hyperfine states are shown
in Fig.\,\ref{fig:sBoundspectrum}(a). The $^{133}$Cs scattering curve is from
the supplemental material of \cite{Berninger2013} and the $^{87}$Rb scattering
curve was calculated using {\sc MOLSCAT} \cite{molscat:v14} with the same
methods outlined in \cite{Blackley:85Rb:2013}, adjusted for the $^{87}$Rb
parameters.

The interspecies \emph{s}-wave scattering length in the lowest spin channel of
$^{87}$Rb + $^{133}$Cs in the experimentally accessible field region is shown
in the top panel of Fig.\,\ref{fig:sBoundspectrum}\,(b). The resonance
positions and widths were calculated using MOLSCAT \cite{molscat:v14} as
modified to handle collisions in magnetic fields \cite{Hutson:field:2011},
using the same basis set and methods as outlined in \cite{Takekoshi:RbCs:2012}.
The parameters of the interatomic potentials were set by fitting the calculated
Feshbach spectrum to experimental measurements in the field range of 0 to
600\,G \cite{Pilch:2009,Takekoshi:RbCs:2012}. The background scattering length
is $+651(\pm10)\,a_0$ associated with the existence of a least-bound state with
a binding energy of 110(2)\,kHz\,$\times h$ \cite{Takekoshi:RbCs:2012}. The
bound states immediately below threshold ($<2$~GHz $\times h$) which are shown
in the bottom panel of Fig.\,\ref{fig:sBoundspectrum}(b) were also obtained in
the manner outlined in \cite{Takekoshi:RbCs:2012}, using the BOUND package
\cite{Hutson:bound:1993}. For the purposes of this study, all states with
physical end-over-end rotation ($L>0$) were excluded to make the figure and
interpretation less cluttered. As in \cite{Takekoshi:RbCs:2012}, bound states
are labelled as $|n(f_{\rm Rb},f_{\rm Cs})L(m_{f_{\rm Rb}},m_{f_{\rm
Cs}})\rangle$, where $n$ is the vibrational label for the particular hyperfine
$(f_{\rm Rb},f_{\rm Cs})$ manifold counting down from the least-bound state
which has $n=-1$.  One additional quantum number, $M_{\rm tot}$, is omitted
because in this study its value is always 4.

The high positive value of the background interspecies atomic scattering length
determines the characteristics of the ultracold mixture. It leads to large
three-body losses during the cooling process, which makes it difficult to
produce a large sample of ultracold atoms at high phase-space densities
suitable for magnetoassociation \cite{Lercher:2011,McCarron:2011,Cho2011}. When
quantum degeneracy is reached it leads to a phase separation of the two
condensates \cite{Pattinson2013}. Fortunately the interspecies scattering
length shows many Feshbach resonances as well. The broad resonances due to $s$-wave states
offer the possibility of improving the evaporation efficiency through tuning
the interspecies collision cross section. In this context the two previously
unobserved high-field resonances at 790 and 1115\,G are interesting candidates.
There are also many narrower resonances on which to explore magnetoassociation
of molecules.

\subsection{Experimental details}

\begin{figure}%
\includegraphics[width=0.8\columnwidth]{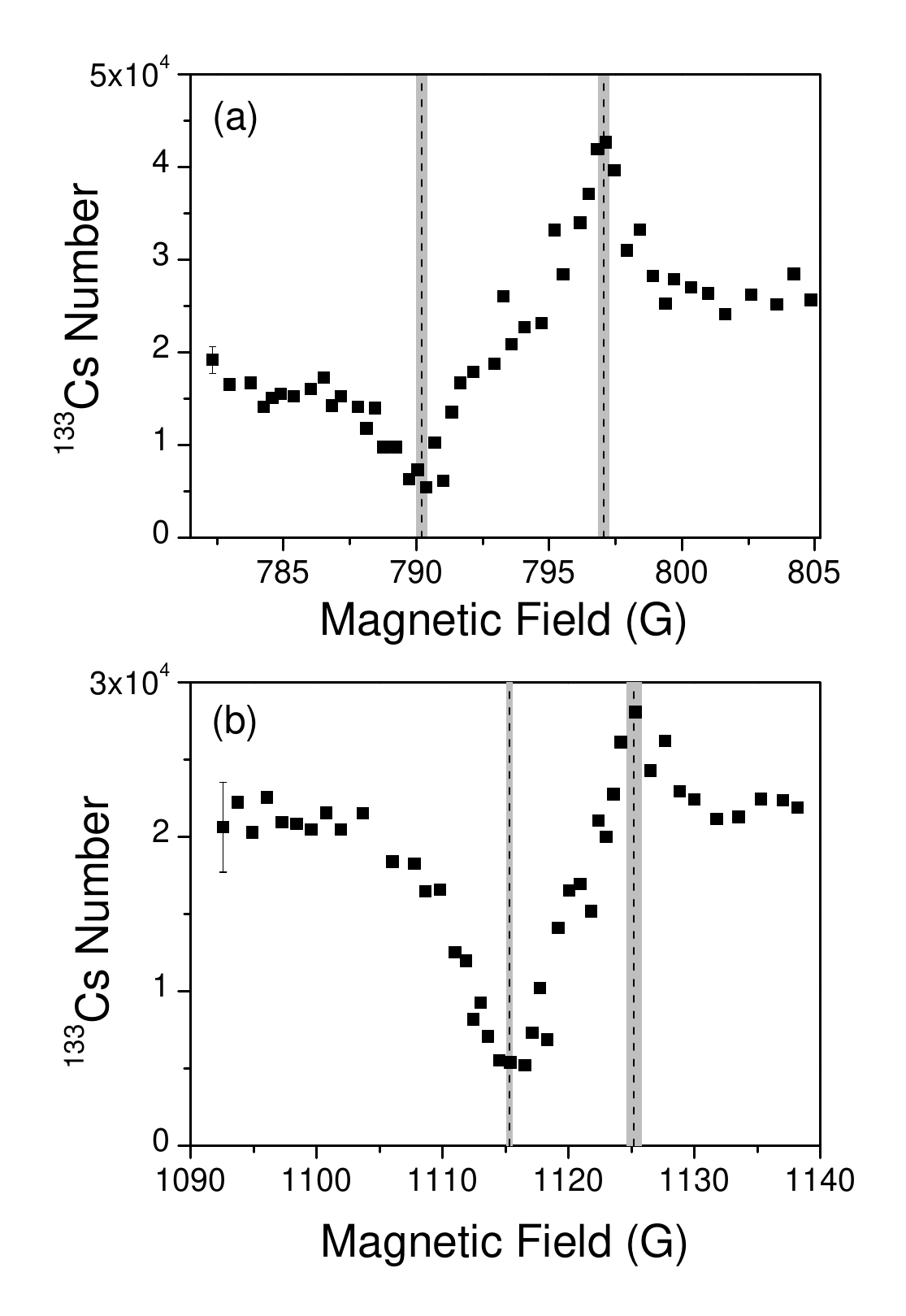}
\caption{Interspecies Feshbach resonances measured at (a) 790.2(2)\,G and (b)
1115.2(2)\,G. The $^{133}$Cs atom number shows a minimum and a maximum for each
resonance. The minima define the resonance positions, while experimental widths
are determined by the difference between the positions of the minima and
maxima. Lorentzian fits to specific ranges of the results determine the
positions. Gray shaded regions indicate the uncertainty in each position. Error
bars show the standard deviations for multiple control shots at specific
magnetic fields.}
\label{fig:FbResonances}%
\end{figure}

Details of our apparatus have been previously described in the context of
studies on dual-species $^{87}$Rb--$^{133}$Cs condensates
\cite{McCarron:2011,Cho2011} and Feshbach spectroscopy of an $^{85}$Rb and
$^{133}$Cs mixture \cite{Cho2013}. Initially cold atoms are collected in a
two-species magneto-optical trap. The $^{87}$Rb and $^{133}$Cs atoms are then
optically pumped into the $(f=1,m_{f}=-1)$ and $(3,-3)$ states, respectively
and loaded into a magnetic quadrupole trap. The $^{87}$Rb atoms are then
further cooled by forced RF evaporation while interspecies elastic collisions
cool the $^{133}$Cs sympathetically. At $40\,\mu$K further efficient
evaporation is prevented by Majorana losses \cite{Lin:hybrid:2009} and the two
species are transferred into a crossed-beam dipole trap formed using a 30\,W
single-frequency fibre laser at 1550\,nm. The crossed trap consists of two
independent beams, typically up to 6~W in power, focussed to waists of
70\,$\mu$m. The beams intersect at an angle of 22$^\circ$, at a position
$\sim$100\,$\mu$m below the field zero of the quadrupole trap. The dipole trap
is loaded by simply reducing the magnetic field gradient of the quadrupole trap
to 29\,G/cm. Immediately following the transfer to the dipole trap, a 22\,G
bias field in the vertical direction is ramped on in 18\,ms and RF adiabatic
rapid passage is used to transfer the $^{87}$Rb and $^{133}$Cs atoms into the
$(1,+1)$ and $(3,+3)$ states respectively \cite{Jenkin:2011}. The direction of
the bias field is chosen such that the final high-field-seeking states of the
atoms are levitated against gravity. The magnetic field gradient of 29\,G/cm is
chosen to be just below the 30.1\,G/cm (30.5\,G/cm) required to levitate
$^{87}$Rb ($^{133}$Cs) at this magnetic field in order to aid evaporation from
the trap. The near-identical ratios of magnetic moment to mass for this mixture
ensure excellent spatial overlap of the two species in the trap. The dipole
trap depth for $^{133}$Cs is  $\sim1.4$ times deeper than that for $^{87}$Rb
due to the different polarizabilities at 1550\,nm, making the trap well suited
for sympathetic cooling of $^{133}$Cs via the direct evaporation of $^{87}$Rb
by decreasing the power of the dipole trap beams. The large interspecies
scattering length ensures that the two species remain in thermal equilibrium.
To probe the mixture the trap and magnetic fields are switched off and resonant
absorption images are captured for both species simultaneously using a
frame-transfer CCD camera \cite{Harris:2008}.

\begin{table*}
\begin{center}
\begin{tabular}{llcccccccccc}
\hline
\hline
 \multicolumn{5}{c} {Experiment}	&& \multicolumn{3}{c} {Theory}\\
 \cline{1-5}
 \cline{7-9}
 \multicolumn{2}{c} {This work} && \multicolumn{2}{c} {Previous work\cite{Takekoshi:RbCs:2012}}\\
 \cline{1-2}
 \cline{4-5}
 $B_{0}$ (G) & $\Delta_{\rm{exp}}$(G) &	&  $B_{0}$ (G) & $\Delta_{\rm{exp}}$(G) & & Quantum labels & $B_{0}$ (G) & $\Delta$ (G) \\
 \hline
 \multicolumn{9}{c} {Resonances due to $s$-wave states}\\
279.03(1)  & 0.11(1)  &&	279.12(5)	&	0.09(3)	&&	$\left|-6(2,4)s(2,2)\right\rangle$ & 279.020 & 0.034 \\
310.72(2)  & 0.70(3)  &&	310.69(6)	&	0.60(4)	&&	$\left|-6(2,4)s(1,3)\right\rangle$ & 310.714 & 0.586 \\
352.7(2)  & 2.9(5)  &&	352.65(34)	&	2.70(47)	&&	$\left|-6(2,4)s(0,4)\right\rangle$ & 352.744 & 2.218 \\
790.2(2)  & 6.8(2)  &&			&			&&	$\left|-5(2,3)s(2,2)\right\rangle$ & 791.791 & 4.227 \\
1115.2(2) & 10.0(6) &&			&			&&	$\left|-5(2,3)s(1,3)\right\rangle$ & 1116.554 & 8.954 	\\ \hline
 \multicolumn{9}{c} {Resonances due to $d$-wave states} \\
181.55(5)	& \hspace{3mm}--	 &&	181.64(8)	&	0.27(10)	&&	$\ket{-6(2,4)d(2,4)}$ & 181.631	&	0.183	\\
197.10(3)	&	0.1(1)	&&	197.06(5)	&	0.09(1)	&&	$\ket{-6(2,4)d(2,3)}$	&	197.065	&	0.052	\\
910.6(8)	&	\hspace{3mm}--		&&	&	&&	$\left|-5(2,3)d(1,2)\right\rangle$	&	909.345	&	0.006	\\
\hline
\hline
\end{tabular}
\end{center}
\caption{Interspecies Feshbach resonances between $^{87}$Rb $(1,+1)$ and
$^{133}$Cs $(3,+3)$ atoms. All five resonances resulting from $s$-wave states shown in the bound-state picture in Fig.\,\ref{fig:sBoundspectrum}\,(b) are reported.
In addition, we list a previously unobserved resonance due to a $d$-wave state at high
magnetic field, together with measurements of the two resonances relevant for
the magnetoassociation results presented in Sec.\,\ref{sec:Molecules}. This
work's experimental errors shown are statistical standard errors resulting from
fits as described in the text. Additional systematic uncertainties of 0.1\,G
and 0.5\,G apply to resonance positions in the field ranges 0 to 400\,G and 400
to 1200\,G respectively. The results of previous measurements are listed for
comparison. Theoretical widths correspond the difference in magnetic field
between the pole and zero in the scattering length.} \label{tab:s-Resonances}
\end{table*}

Typically, $2.8(2)\times10^6$ $^{87}$Rb atoms at a temperature of
9.6(1)$\,\mu$K are confined in the dipole trap following the spin flip. The
number of $^{133}$Cs atoms collected in the magneto-optical trap is actively
controlled allowing the number ratio of $^{87}$Rb and $^{133}$Cs to be varied
precisely. The depth of the trap for $^{87}$Rb ($^{133}$Cs) is then decreased
from 76\,$\mu$K (103\,$\mu$K) to 3.1\,$\mu$K (4.2\,$\mu$K) over 1.25 seconds.
The majority of the evaporation is performed at a bias field of 22\,G where
$^{133}$Cs can be cooled efficiently, though the final stages of cooling are
performed at a magnetic field within the field range under investigation.
Following the evaporation, a sample of typically $\sim 5 \times 10^{5}$
$^{87}$Rb atoms and $\sim 5 \times 10^{4}$ $^{133}$Cs atoms with a temperature
between 200 and 700\,nK remains confined in the dipole trap. Feshbach
spectroscopy is then performed by exposing the ultracold atomic mixture to
different magnetic bias fields and detecting the atom loss. The hold time at
each field is adjusted to between 300 and 500\,ms for different loss features
to give the clearest signal. The significant imbalance in atom number between
the two species increases the sensitivity of heteronuclear Feshbach
spectroscopy; the $^{133}$Cs atoms act as a probe species immersed in a
collisional bath of $^{87}$Rb atoms \cite{Wille2008}. The magnetic field is
calibrated using microwave spectroscopy between Zeeman components of the
different hyperfine states of $^{133}$Cs. These measurements reveal the
long-term reproducibility of the field to be 0.1\,G in the range 0 to 400\,G
and 0.5\,G in the range 400 to 1200\,G.

\subsection{Results}

The scattering length $a(B)$ passes through a pole at a Feshbach resonance, and
in the vicinity of the resonance may be represented $a(B) =
a_{\rm{bg}}\left[1-\Delta/(B-B_0)\right]$ \cite{Moerdijk:1995}, where
$a_{\rm{bg}}$ is the background scattering length, $B_0$ is the resonance
position and $\Delta$ is the width. The three-body recombination rate scales
approximately as $a^{4}$ \cite{Fedichev:a4:1996}. Typical signatures of the
resonances we observe in three-body loss spectra are shown in Fig.\
\ref{fig:FbResonances}, for the two resonances due to $s$-wave states above 600\,G. In both
cases, the $^{133}$Cs atom number shows a significant drop followed by a
pronounced peak as the resonance is crossed from low to high magnetic fields.
The drop in atom number is due to the increase in the recombination rate near
the pole of the resonance. Fitting a Lorentzian lineshape to the loss feature
allows the resonance position $B_{0}$ to be assigned. Conversely, near the
zero-crossing in the scattering length the three-body loss rate is reduced from
the background value and the efficiency of sympathetic cooling improves,
leading to an observed peak in the atom number. Fitting a Lorentzian lineshape
to the peak in atom number allows an experimental width ($\Delta_{\rm{exp}}$)
to be determined for the resonance, defined to be the magnetic field difference
between the minimum and maximum in the atom number.

In Table \ref{tab:s-Resonances} we report measurements of all five resonances due to $s$-wave states 
below 1200\,G resulting from the bound-state picture shown in
Fig.\,\ref{fig:sBoundspectrum}\,(b). For all of these resonances we can measure
an experimental width ($\Delta_{\rm{exp}}$). We can also calculate the width
$\Delta$ from the positions of the pole and zero-crossing obtained from our
coupled-channels model. Ongoing theoretical work \cite{Wang2013} suggests that
the width obtained from 3-body loss rates is systematically smaller than the
2-body width obtained from the pole and zero-crossing in the scattering length,
particularly for narrow resonances, and the present results appear to support
this conclusion. In addition, we list in Table \ref{tab:s-Resonances} a
previously unobserved resonance due to a $d$-wave state at high magnetic field, together with
the two resonances relevant for the magnetoassociation results presented in
Sec.\,\ref{sec:Molecules}. Where available we also give the results of previous
measurements \cite{Takekoshi:RbCs:2012} for comparison.

\section{Magnetoassociation of $^{87}$R\lowercase{b}C\lowercase{s} Feshbach molecules}\label{sec:Molecules}

\subsection{Overview}

Feshbach molecules may be created by sweeping the applied magnetic field across
a Feshbach resonance such that the energy of the separated atomic states is
tuned adiabatically through an avoided crossing with the energy of a weakly
bound molecular state \cite{Chin:RMP:2010}. Such a magnetoassociation sequence
generally transfers a fraction of the atomic sample into the weakly bound
molecular state responsible for the Feshbach resonance. The efficiency of the
conversion from atoms to molecules is largely determined by the phase-space
density of the atomic gas \cite{Hodby:2005}. Further changes in the magnetic
field allow navigation through the rich spectrum of near-threshold bound states
(see the lower panel of Fig.\,\ref{fig:sBoundspectrum}\,(b) for example),
traversing the avoided crossings between molecular states either diabatically
or adiabatically depending on the rate of change of the magnetic field
\cite{Mark:spect:2007}. The molecules may be separated from the atomic cloud by
exploiting the difference in the magnetic moments of the molecules and the
atoms; time of flight expansion in the presence of an applied magnetic field
gradient leads to a Stern-Gerlach separation of the atoms and molecules
\cite{Herbig:2003}. In bosonic gases, a fast separation is essential to
minimise inelastic collisions between the atoms and molecules which lead to
loss, generally associated with the molecules being transferred to more deeply
bound states. Following the separation, the molecular cloud can be detected by
reversing the association sequence and imaging the atoms that result from the
dissociation of the molecules \cite{Mukaiyama:2004,Durr:diss:2004}.

\subsection{Magneto-association sequence}

\begin{figure}%
\includegraphics[width=1\columnwidth]{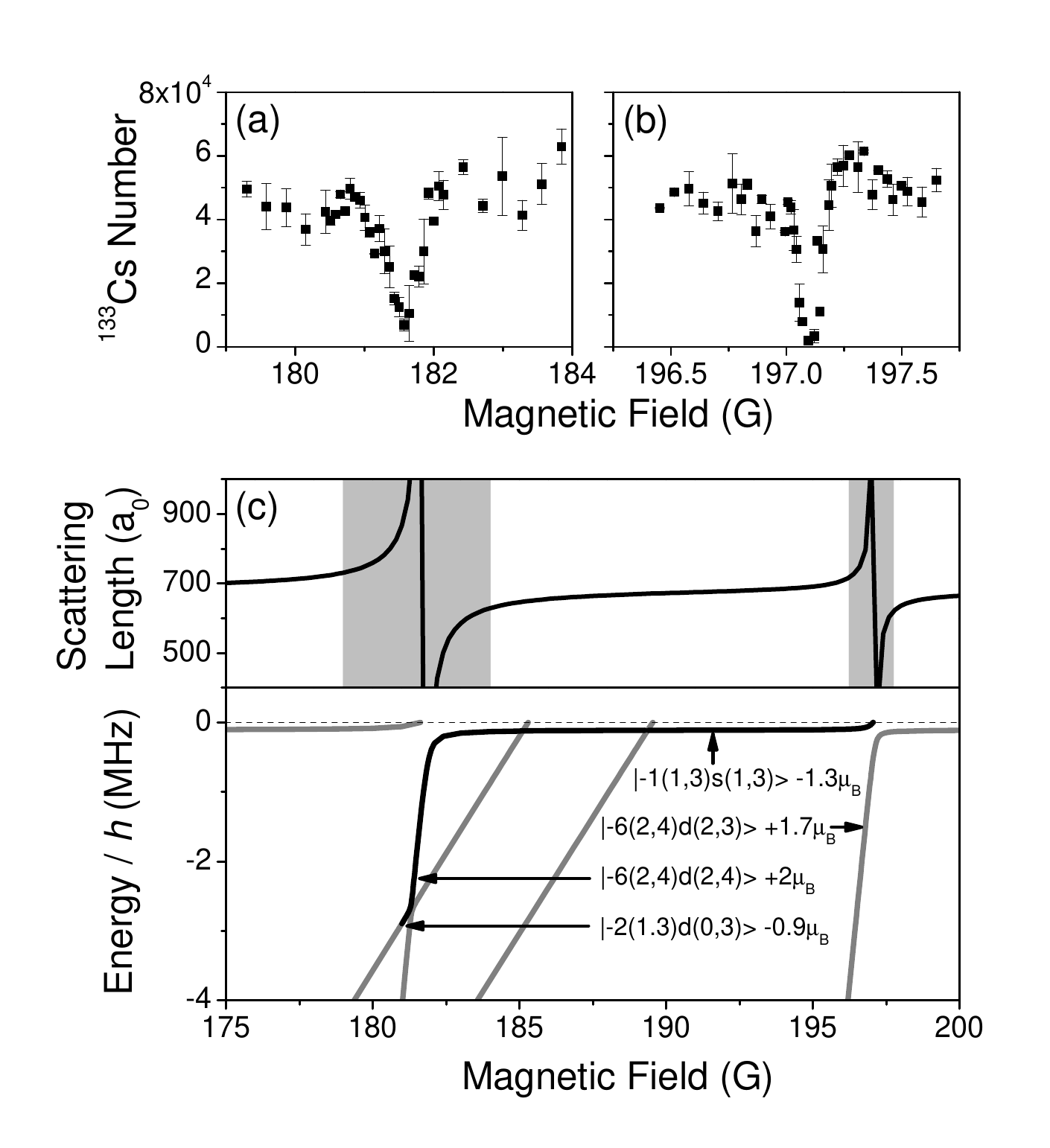}
\caption{The magnetoassociation sequence. The interspecies Feshbach resonances
at (a) 181.55(5)\,G and (b) 197.10(3)\,G detected through loss in the
$^{133}$Cs atom number as described in Sec.\,\ref{sec:Resonances}. (c) Upper
panel: The interspecies scattering length between $^{87}$Rb and $^{133}$Cs in
the relevant magnetic field range. The grey shaded areas mark the field ranges
shown in (a) and (b). Lower panel: The weakly bound molecular states relevant
to the magnetoassociation sequence calculated using the $s$ and $d$ basis
functions as discussed in the text. Also shown are the magnetic moments for
each bound state. Molecules are produced at the Feshbach resonance at
197.10(3)\,G and then transferred into the $\ket{-2(1,3)d(0,3)}$ state at
181\,G following the path shown by the solid black line.}
\label{fig:BoundState}%
\end{figure}

The bound-state spectrum relevant to magnetoassociation of $^{87}$RbCs
molecules is illustrated in the lower panel of Fig.\,\ref{fig:BoundState}\,(c).
The creation and separation of molecules is complicated by the presence of the
$\ket{-1(1,3)s(1,3)}$ bound state, which runs parallel to the atomic threshold
at a binding energy of 110(2)\,kHz\,$\times h$ \cite{Takekoshi:RbCs:2012}. This
state leads to strong avoided crossings just below threshold as the more deeply
bound molecular states responsible for the Feshbach resonances approach
threshold. As a consequence sweeping across a Feshbach resonance from high to
low field creates molecules in the near-threshold $\ket{-1(1,3)s(1,3)}$ state
with a magnetic moment of $-1.3\,\mu_{\rm{B}}$. Crucially, the ratio of
magnetic moment to mass for molecules in this state is almost identical to that
for $^{87}$Rb in the (1,+1) state \emph{and} $^{133}$Cs in the (3,+3) state.
Consequently, in order for the Stern-Gerlach separation to work, the molecules
must be transferred into a different state with a substantially different
magnetic moment. We therefore associate the molecules on the Feshbach resonance
arising from the $\ket{-6(2,4)d(2,3)}$ state at 197.10(3)\,G and then sweep the
magnetic field to 181\,G, adiabatically following the avoided crossing below
the Feshbach resonance at 181.55(5)\,G and transferring the molecules via the
weak-field-seeking $\ket{-6(2,4)d(2,4)}$ state into the high-field-seeking
$\ket{-2(1,3)d(0,3)}$ state with a magnetic moment of $-0.9\,\mu_{\rm{B}}$.
This path is illustrated in the lower panel of Fig.\,\ref{fig:BoundState}\,(c)
by the solid black line.

\begin{figure}%
\includegraphics[width=1\columnwidth]{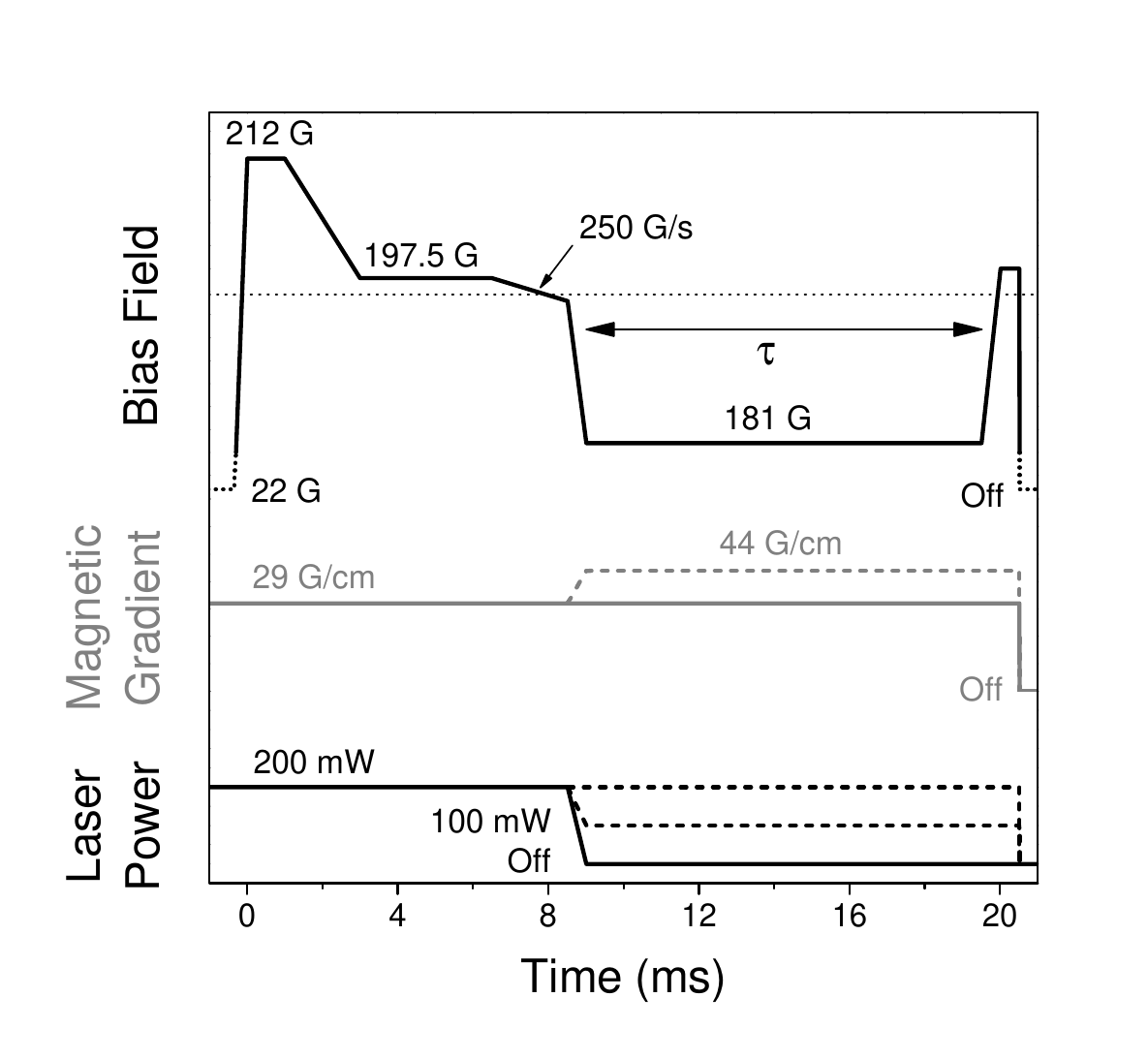}
\caption{The experimental sequence for the creation, Stern-Gerlach separation
and detection of Feshbach molecules. An ultracold atomic mixture is initially
created by evaporation in the dipole trap at a bias field of 22\,G.
Subsequently the timing sequences shown for the bias field, magnetic field
gradient and laser power in each beam of the crossed dipole trap are applied.
The horizontal dotted line indicates the position of the Feshbach resonance at
197.10(3)\,G used for magnetoassociation. The dashed lines show the changes to
routine implemented to trap the atoms in the dipole trap. The hold time,
$\tau$, in the dipole trap is varied to measure the lifetime of the molecules.}
\label{fig:Association}%
\end{figure}

The timing of the magnetoassociation sequence is shown in
Fig.\,\ref{fig:Association}. An ultracold atomic mixture is produced by
evaporation in the dipole trap at a bias field of 22\,G. The magnetic field is
then quickly increased to 212\,G to avoid three-body losses induced by crossing
several interspecies and intraspecies Feshbach resonances (see
Fig.\,\ref{fig:sBoundspectrum}). From here the field is decreased to 197.5\,G,
very close to the upper end of the resonance loss feature (see
Fig.\,\ref{fig:BoundState}\,(b)). After a hold time of 3.5\,ms to allow the
magnetic field to settle, the field is further decreased by $\sim0.5$\,G in
2\,ms. During this ramp, molecules in the least-bound state
$\ket{-1(1,3)s(1,3)}$ are produced. In the current setup, the magnetic coils
can levitate only high-field-seeking states against gravity. Therefore to
separate the molecules and atoms we subsequently switch the field to 181\,G in
0.1\,ms, thereby transferring the molecules into $\ket{-2(1,3)d(0,3)}$ state.
At the same time we switch off the optical dipole trap. After 7~ms time of
flight, the atoms and molecules are completely separated. To detect the
molecules, the magnetic field is ramped from 181\,G to 199.7\,G in 0.5\,ms,
reversing the association sequence and dissociating the molecular cloud. The
magnetic bias field and gradient are then switched off and resonant absorption
imaging is used to image both $^{87}$Rb and $^{133}$Cs onto a CCD camera.

\begin{figure}%
\includegraphics[width=1\columnwidth]{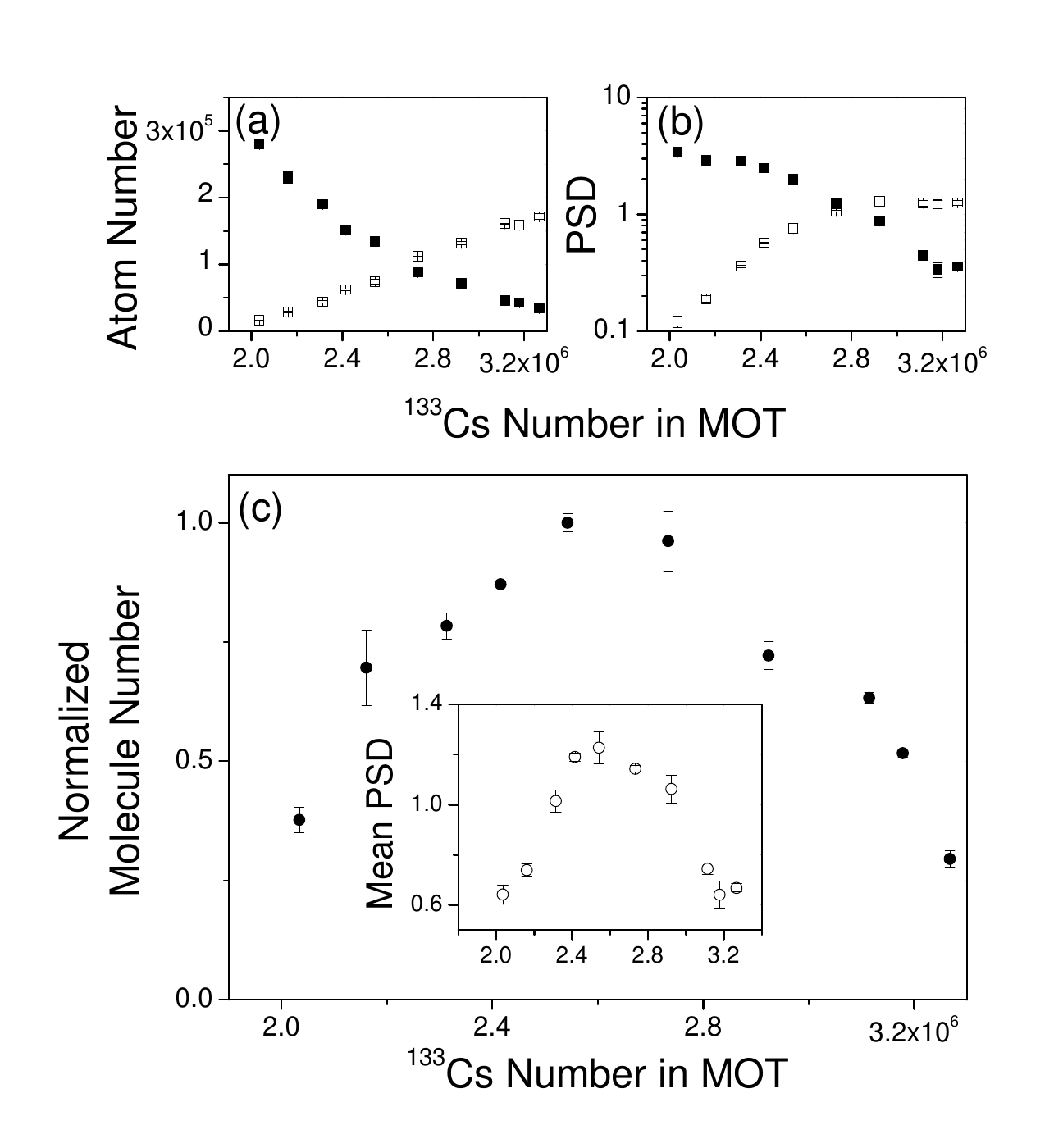}
\caption{Optimising the production of Feshbach molecules by changing the
composition of the initial ultracold atomic mixture. (a) Atom number and (b)
phase-space density for $^{87}$Rb (closed squares) and $^{133}$Cs (open
squares) at the end of a fixed evaporation sequence as a function of the number
of $^{133}$Cs atoms loaded into the magneto-optical trap. (c) The corresponding
number of molecules produced by magnetoassociation normalised to the peak
number ($\sim3000$ molecules in this case). The inset shows the geometric mean
of the $^{87}$Rb and $^{133}$Cs phase-space densities. A strong correlation
between the molecule conversion efficiency and the mean phase-space density is
observed.}
\label{fig:ratio}%
\end{figure}

\subsection{Optimising the molecule production}

The parameters of the ramp sequence reported above were all carefully optimised
to maximise the number of molecules produced. The most sensitive parameter
affecting the molecule production proved to be the composition of the initial
ultracold atomic mixture. To explore this systematically, we varied the ratio
of $^{87}$Rb and $^{133}$Cs by changing the number of $^{133}$Cs atoms loaded
into the MOT. The results are shown in Fig.\,\ref{fig:ratio}. The performance
of the evaporation sequence is extremely sensitive to the initial number of
$^{133}$Cs atoms loaded into the MOT. Figs.\,\ref{fig:ratio}\,(a) and (b) show
that, for a fixed evaporation ramp, a 50\,\% change in the $^{133}$Cs atom
number results in an order-of-magnitude change in the number and phase-space
density of both $^{87}$Rb and $^{133}$Cs. This follows from our reliance on
sympathetic cooling of $^{133}$Cs throughout the evaporation sequence.
Concomitantly there is a sharp variation in the number of $^{87}$RbCs molecules
produced, as shown in Fig.\,\ref{fig:ratio}\,(c). The molecular conversion
efficiency closely follows the geometric mean of the $^{87}$Rb and $^{133}$Cs
phase-space densities, as shown in the inset of Fig.\,\ref{fig:ratio}\,(c). At
the peak of the plot we produce about 3000 molecules from an ultracold mixture
containing $\sim1\times10^5$ atoms of each species at a temperature of
0.1\,$\mu$K. Changing the end point of the evaporation and adjusting the number
of $^{133}$Cs atoms loaded into the MOT to maximise the average phase-space
density leads to the production of slightly more molecules. Under optimum
conditions we observe up to $\sim5000$ molecules produced from a sample of $2
\times 10^5$ atoms of each species at a temperature of 0.3~$\mu$K. This
represents a conversion efficiency of $\sim2.5$\%.

\subsection{Towards ground state molecules}

\begin{figure}%
\includegraphics[width=1\columnwidth]{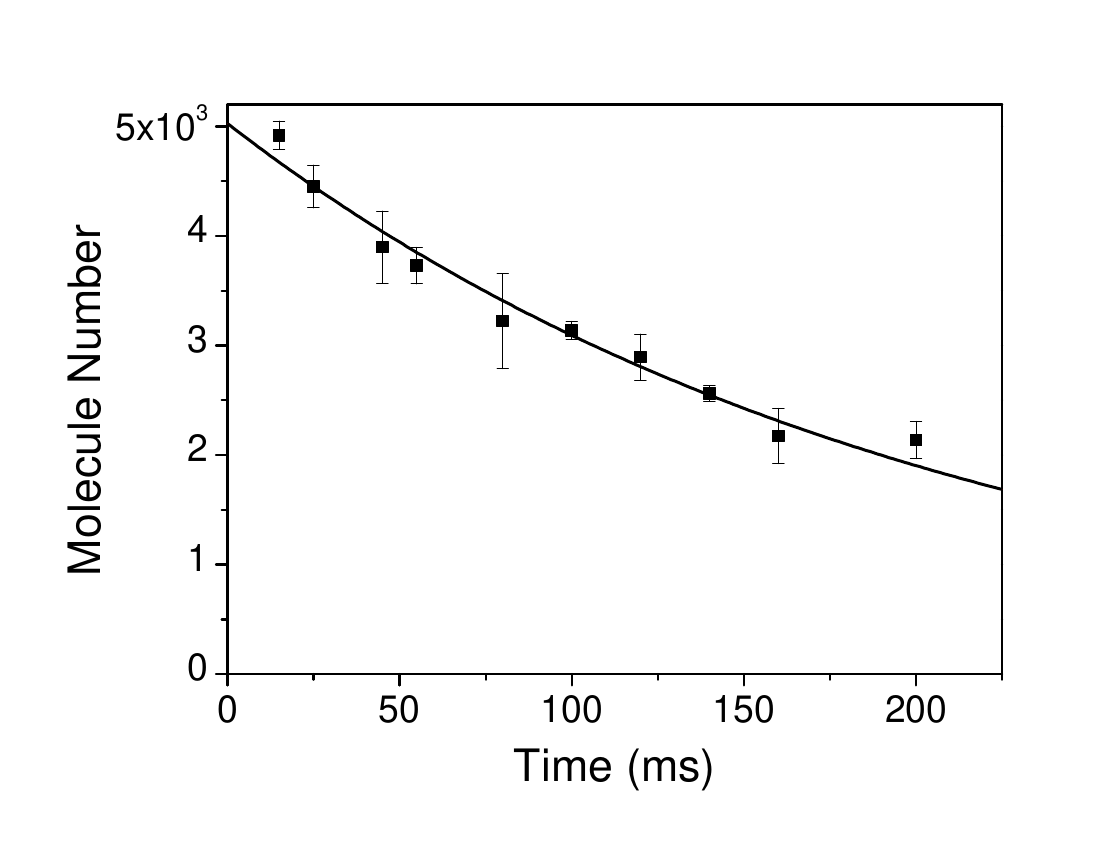}
\caption{Lifetime of $^{87}$RbCs molecules in the $\ket{-2(1,3)d(0,3)}$ state
at 181\,G held in the optical dipole trap. A magnetic field gradient of
44\,G/cm is applied to levitate the molecules and the power in each beam of the
dipole trap is 100\,mW. Under these conditions there is no trapping potential
for the atoms. The solid line shows an exponential fit to the results, which
yields a lifetime of 0.21(1)\,s.}
\label{fig:TrappingTime}%
\end{figure}

For future experiments exploring the optical transfer of the molecules to the
rovibrational ground state of the singlet potential, it is desirable to confine
the Feshbach molecules in the optical dipole trap. This is achieved using a
slightly different experimental routine shown by the dashed lines in
Fig.\,\ref{fig:Association}. Now the power in the dipole trap remains on at a
variable level and the magnetic field gradient is increased to 44\,G/cm to
levitate the molecules in the $\ket{-2(1,3)d(0,3)}$ state. This gradient
over-levitates the atoms, resulting in a much reduced trap depth. Consequently
all the unconverted atoms escape from the trap within 5\,ms leaving a pure
sample of molecules. To measure the lifetime of the molecules in the trap, we
also reduce the power in each beam of the dipole trap to 100\,mW. Under these
conditions there is no trapping potential for the atoms, so that our lifetime
measurements are not contaminated by residual atomic signals. The results are
shown in Fig.\,\ref{fig:TrappingTime}. We observe a lifetime of the Feshbach
molecules in the $\ket{-2(1,3)d(0,3)}$ state of 0.21(1)\,ms at a trapped
density of $\sim8\times10^9$\,cm$^{-3}$.

\begin{figure}%
\includegraphics[width=1\columnwidth]{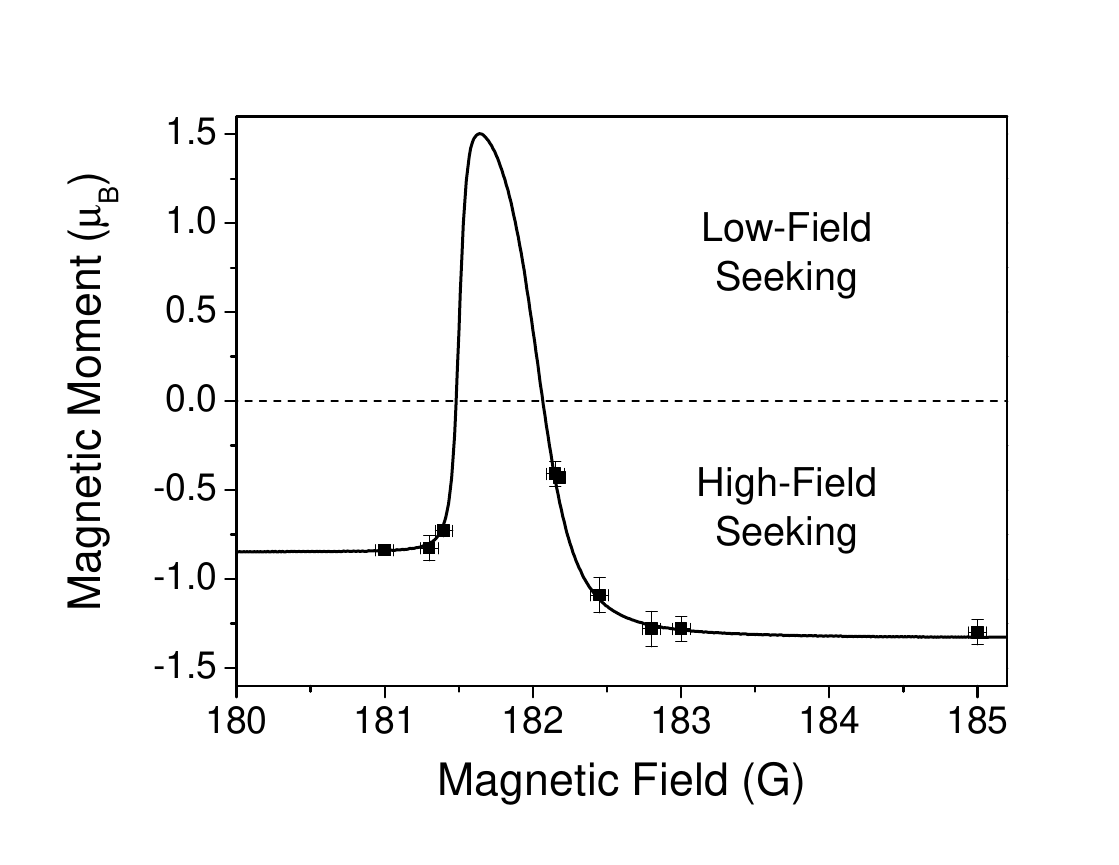}
\caption{Magnetic moment of the $^{87}$RbCs molecules as a function of the
magnetic bias field. The experimental points are determined by finding the
magnetic field gradient at which the molecules are levitated. Note the
measurements are restricted to magnetic moments in the high-field-seeking
region due to the current coil geometry in the experiment. The solid line is
the theoretical prediction for the molecular states following the
magnetoassociation path shown in Fig.\,\ref{fig:BoundState}\,(c).}
\label{fig:MagnMoment}%
\end{figure}

We have also measured the magnetic moment of the molecules as a function of the
bias field. For this measurement Feshbach molecules in the
$\ket{-2(1,3)d(0,3)}$ state are first loaded into the dipole trap as above.
After a hold time of 50\,ms the trap is switched off and the magnetic bias
field and gradient are switched to different values. Measuring the magnetic
field gradient at which the sample is levitated against gravity allows the
magnetic moment of the molecules to be determined. The results are shown in
Fig.\,\ref{fig:MagnMoment} for the magnetic bias-field region between 180 and
185\,G. Note that with the current coil configuration we can measure the
magnetic moment of the molecules only whilst they are in a high-field-seeking
state. The results are compared to magnetic moment calculated from the
bound-state picture shown in the lower panel of
Fig.\,\ref{fig:BoundState}\,(c). In this plot the zero-energy threshold
corresponds to the energy of the two unbound atoms, namely $^{87}$Rb in the
(1,+1) state and $^{133}$Cs in the (3,+3) state. This energy varies with
magnetic field according to the usual Breit-Rabi formula for each atom. The
slope of the energy of bound states with magnetic field, taking into account
the Zeeman shift of the unbound atoms, then gives the magnetic moment of the
molecules. The result for the magnetoassociation path illustrated by the solid
black line in Fig.\,\ref{fig:BoundState}\,(c) is shown by the solid line in
Fig.\,\ref{fig:MagnMoment}. There is excellent agreement between theory and
experiment. These results also demonstrate the ability to control the character
of the molecular state. Such control can be used to improve the overlap between
the Feshbach molecule and electronically excited states suitable for use in
optical transfer schemes to the rovibrational ground state
\cite{Danzl:v73:2008}.

\section{Conclusion and Outlook}\label{sec:Conclusion}

In summary we have produced ultracold samples of up to $\sim5000$ $^{87}$RbCs
Feshbach molecules using magnetoassociation on an interspecies Feshbach
resonance. The molecules are formed from an atomic sample with a temperature of
300\,nK and can be confined in an optical dipole trap with a lifetime of
0.21(1)\,s. We have demonstrated the control of the character of the molecules
by adiabatic following of the bound-state spectrum using magnetic field ramps.
The overall magnetoassociation efficiency observed in our experiment is in good
agreement with results reported by Takekoshi \emph{et al.}\
\cite{Takekoshi:RbCs:2012}, despite very different approaches used in producing
the initial ultracold atomic mixture. In the course of this work, we have also
detected several previously unobserved interspecies Feshbach resonances in the
high magnetic field region. The positions of these resonances agree very well
with the theoretical predictions, further supporting the quality of the fitted
interspecies potential \cite{Takekoshi:RbCs:2012}.

These results represent a good starting point for exploring the optical
transfer from the Feshbach state to the rovibrational singlet ground state
using stimulated Raman adiabatic passage
\cite{Bergmann:1998,Ni:KRb:2008,Lang:ground:2008,Danzl:ground:2010}. The
relevant molecular spectroscopy for ground state transfer of $^{87}$RbCs has
recently been reported by Debatin \emph{et al.}\ \cite{Debatin:2011}. We have
constructed a suitable laser system in which the two lasers used in the optical
transfer are frequency stabilised to a common ultra-low expansion optical
resonator \cite{Aikawa2011} and are currently investigating one- and two-photon
optical spectroscopy of the ultracold $^{87}$RbCs Feshbach molecules. The
subsequent transfer to the ground state will realise an ultracold gas of stable
polar molecules with numerous applications.

\section{Acknowledgements}

We acknowledge useful discussions with H.-C. N\"{a}gerl and members of his
group and with P. S. Julienne. This work was supported by the UK EPSRC and by
EOARD Grant FA8655-10-1-3033. CLB is supported by a Doctoral Fellowship from
Durham University.


\begin{thebibliography}{59}%
\makeatletter
\providecommand \@ifxundefined [1]{%
 \@ifx{#1\undefined}
}%
\providecommand \@ifnum [1]{%
 \ifnum #1\expandafter \@firstoftwo
 \else \expandafter \@secondoftwo
 \fi
}%
\providecommand \@ifx [1]{%
 \ifx #1\expandafter \@firstoftwo
 \else \expandafter \@secondoftwo
 \fi
}%
\providecommand \natexlab [1]{#1}%
\providecommand \enquote  [1]{``#1''}%
\providecommand \bibnamefont  [1]{#1}%
\providecommand \bibfnamefont [1]{#1}%
\providecommand \citenamefont [1]{#1}%
\providecommand \href@noop [0]{\@secondoftwo}%
\providecommand \href [0]{\begingroup \@sanitize@url \@href}%
\providecommand \@href[1]{\@@startlink{#1}\@@href}%
\providecommand \@@href[1]{\endgroup#1\@@endlink}%
\providecommand \@sanitize@url [0]{\catcode `\\12\catcode `\$12\catcode
  `\&12\catcode `\#12\catcode `\^12\catcode `\_12\catcode `\%12\relax}%
\providecommand \@@startlink[1]{}%
\providecommand \@@endlink[0]{}%
\providecommand \url  [0]{\begingroup\@sanitize@url \@url }%
\providecommand \@url [1]{\endgroup\@href {#1}{\urlprefix }}%
\providecommand \urlprefix  [0]{URL }%
\providecommand \Eprint [0]{\href }%
\providecommand \doibase [0]{http://dx.doi.org/}%
\providecommand \selectlanguage [0]{\@gobble}%
\providecommand \bibinfo  [0]{\@secondoftwo}%
\providecommand \bibfield  [0]{\@secondoftwo}%
\providecommand \translation [1]{[#1]}%
\providecommand \BibitemOpen [0]{}%
\providecommand \bibitemStop [0]{}%
\providecommand \bibitemNoStop [0]{.\EOS\space}%
\providecommand \EOS [0]{\spacefactor3000\relax}%
\providecommand \BibitemShut  [1]{\csname bibitem#1\endcsname}%
\let\auto@bib@innerbib\@empty
\bibitem [{\citenamefont {Carr}\ \emph {et~al.}(2009)\citenamefont {Carr},
  \citenamefont {{DeMille}}, \citenamefont {Krems},\ and\ \citenamefont
  {Ye}}]{Carr:NJPintro:2009}%
  \BibitemOpen
  \bibfield  {author} {\bibinfo {author} {\bibfnamefont {L.~D.}\ \bibnamefont
  {Carr}}, \bibinfo {author} {\bibfnamefont {D.}~\bibnamefont {{DeMille}}},
  \bibinfo {author} {\bibfnamefont {R.~V.}\ \bibnamefont {Krems}}, \ and\
  \bibinfo {author} {\bibfnamefont {J.}~\bibnamefont {Ye}},\ }\href@noop {}
  {\bibfield  {journal} {\bibinfo  {journal} {NJP}\ }\textbf {\bibinfo {volume}
  {11}},\ \bibinfo {pages} {055049} (\bibinfo {year} {2009})}\BibitemShut
  {NoStop}%
\bibitem [{\citenamefont {Micheli}\ \emph {et~al.}(2006)\citenamefont
  {Micheli}, \citenamefont {Brennen},\ and\ \citenamefont
  {Zoller}}]{Micheli:2006}%
  \BibitemOpen
  \bibfield  {author} {\bibinfo {author} {\bibfnamefont {A.}~\bibnamefont
  {Micheli}}, \bibinfo {author} {\bibfnamefont {G.~K.}\ \bibnamefont
  {Brennen}}, \ and\ \bibinfo {author} {\bibfnamefont {P.}~\bibnamefont
  {Zoller}},\ }\href@noop {} {\bibfield  {journal} {\bibinfo  {journal} {Nature
  Phys.}\ }\textbf {\bibinfo {volume} {2}},\ \bibinfo {pages} {341} (\bibinfo
  {year} {2006})}\BibitemShut {NoStop}%
\bibitem [{\citenamefont {Barnett}\ \emph {et~al.}(2006)\citenamefont
  {Barnett}, \citenamefont {Petrov}, \citenamefont {Lukin},\ and\ \citenamefont
  {Demler}}]{Barnett:2006}%
  \BibitemOpen
  \bibfield  {author} {\bibinfo {author} {\bibfnamefont {R.}~\bibnamefont
  {Barnett}}, \bibinfo {author} {\bibfnamefont {D.}~\bibnamefont {Petrov}},
  \bibinfo {author} {\bibfnamefont {M.}~\bibnamefont {Lukin}}, \ and\ \bibinfo
  {author} {\bibfnamefont {E.}~\bibnamefont {Demler}},\ }\href {\doibase
  10.1103/PhysRevLett.96.190401} {\bibfield  {journal} {\bibinfo  {journal}
  {Phys. Rev. Lett.}\ }\textbf {\bibinfo {volume} {96}},\ \bibinfo {pages}
  {190401} (\bibinfo {year} {2006})}\BibitemShut {NoStop}%
\bibitem [{\citenamefont {DeMille}(2002)}]{DeMille:2002}%
  \BibitemOpen
  \bibfield  {author} {\bibinfo {author} {\bibfnamefont {D.}~\bibnamefont
  {DeMille}},\ }\href@noop {} {\bibfield  {journal} {\bibinfo  {journal} {Phys.
  Rev. Lett.}\ }\textbf {\bibinfo {volume} {88}},\ \bibinfo {pages} {067901}
  (\bibinfo {year} {2002})}\BibitemShut {NoStop}%
\bibitem [{\citenamefont {Andr\'e}\ \emph {et~al.}(2006)\citenamefont
  {Andr\'e}, \citenamefont {DeMille}, \citenamefont {Doyle}, \citenamefont
  {Lukin}, \citenamefont {Maxwell}, \citenamefont {Rabl}, \citenamefont
  {Schoelkopf},\ and\ \citenamefont {Zoller}}]{Andre:2006}%
  \BibitemOpen
  \bibfield  {author} {\bibinfo {author} {\bibfnamefont {A.}~\bibnamefont
  {Andr\'e}}, \bibinfo {author} {\bibfnamefont {D.}~\bibnamefont {DeMille}},
  \bibinfo {author} {\bibfnamefont {J.~M.}\ \bibnamefont {Doyle}}, \bibinfo
  {author} {\bibfnamefont {M.~D.}\ \bibnamefont {Lukin}}, \bibinfo {author}
  {\bibfnamefont {S.~E.}\ \bibnamefont {Maxwell}}, \bibinfo {author}
  {\bibfnamefont {P.}~\bibnamefont {Rabl}}, \bibinfo {author} {\bibfnamefont
  {R.~J.}\ \bibnamefont {Schoelkopf}}, \ and\ \bibinfo {author} {\bibfnamefont
  {P.}~\bibnamefont {Zoller}},\ }\href@noop {} {\bibfield  {journal} {\bibinfo
  {journal} {Nature Phys.}\ }\textbf {\bibinfo {volume} {2}},\ \bibinfo {pages}
  {636} (\bibinfo {year} {2006})}\BibitemShut {NoStop}%
\bibitem [{\citenamefont {Flambaum}\ and\ \citenamefont
  {Kozlov}(2007)}]{Flambaum:2007}%
  \BibitemOpen
  \bibfield  {author} {\bibinfo {author} {\bibfnamefont {V.~V.}\ \bibnamefont
  {Flambaum}}\ and\ \bibinfo {author} {\bibfnamefont {M.~G.}\ \bibnamefont
  {Kozlov}},\ }\href@noop {} {\bibfield  {journal} {\bibinfo  {journal} {Phys.
  Rev. Lett.}\ }\textbf {\bibinfo {volume} {99}},\ \bibinfo {pages} {150801}
  (\bibinfo {year} {2007})}\BibitemShut {NoStop}%
\bibitem [{\citenamefont {Isaev}\ \emph {et~al.}(2010)\citenamefont {Isaev},
  \citenamefont {Hoekstra},\ and\ \citenamefont {Berger}}]{Isaev:2010}%
  \BibitemOpen
  \bibfield  {author} {\bibinfo {author} {\bibfnamefont {T.~A.}\ \bibnamefont
  {Isaev}}, \bibinfo {author} {\bibfnamefont {S.}~\bibnamefont {Hoekstra}}, \
  and\ \bibinfo {author} {\bibfnamefont {R.}~\bibnamefont {Berger}},\
  }\href@noop {} {\bibfield  {journal} {\bibinfo  {journal} {Phys. Rev. A}\
  }\textbf {\bibinfo {volume} {82}},\ \bibinfo {pages} {052521} (\bibinfo
  {year} {2010})}\BibitemShut {NoStop}%
\bibitem [{\citenamefont {Hudson}\ \emph {et~al.}(2011)\citenamefont {Hudson},
  \citenamefont {Kara}, \citenamefont {Smallman}, \citenamefont {Sauer},
  \citenamefont {Tarbutt},\ and\ \citenamefont {Hinds}}]{Hudson:2011}%
  \BibitemOpen
  \bibfield  {author} {\bibinfo {author} {\bibfnamefont {J.~J.}\ \bibnamefont
  {Hudson}}, \bibinfo {author} {\bibfnamefont {D.~M.}\ \bibnamefont {Kara}},
  \bibinfo {author} {\bibfnamefont {I.~J.}\ \bibnamefont {Smallman}}, \bibinfo
  {author} {\bibfnamefont {B.~E.}\ \bibnamefont {Sauer}}, \bibinfo {author}
  {\bibfnamefont {M.~R.}\ \bibnamefont {Tarbutt}}, \ and\ \bibinfo {author}
  {\bibfnamefont {E.~A.}\ \bibnamefont {Hinds}},\ }\href@noop {} {\bibfield
  {journal} {\bibinfo  {journal} {Nature}\ }\textbf {\bibinfo {volume} {473}},\
  \bibinfo {pages} {493} (\bibinfo {year} {2011})}\BibitemShut {NoStop}%
\bibitem [{\citenamefont {Leanhardt}\ \emph {et~al.}(2011)\citenamefont
  {Leanhardt}, \citenamefont {Bohn}, \citenamefont {Loh}, \citenamefont
  {Maletinsky}, \citenamefont {Meyer}, \citenamefont {Sinclair}, \citenamefont
  {Stutz},\ and\ \citenamefont {Cornell}}]{Leanhardt:2011}%
  \BibitemOpen
  \bibfield  {author} {\bibinfo {author} {\bibfnamefont {A.~E.}\ \bibnamefont
  {Leanhardt}}, \bibinfo {author} {\bibfnamefont {J.~L.}\ \bibnamefont {Bohn}},
  \bibinfo {author} {\bibfnamefont {H.}~\bibnamefont {Loh}}, \bibinfo {author}
  {\bibfnamefont {P.}~\bibnamefont {Maletinsky}}, \bibinfo {author}
  {\bibfnamefont {E.~R.}\ \bibnamefont {Meyer}}, \bibinfo {author}
  {\bibfnamefont {L.~C.}\ \bibnamefont {Sinclair}}, \bibinfo {author}
  {\bibfnamefont {R.~P.}\ \bibnamefont {Stutz}}, \ and\ \bibinfo {author}
  {\bibfnamefont {E.~A.}\ \bibnamefont {Cornell}},\ }\href@noop {} {\bibfield
  {journal} {\bibinfo  {journal} {J. Mol. Spectrosc.}\ }\textbf {\bibinfo
  {volume} {270}},\ \bibinfo {pages} {1} (\bibinfo {year} {2011})}\BibitemShut
  {NoStop}%
\bibitem [{\citenamefont {Baron}\ \emph {et~al.}(2013)\citenamefont {Baron},
  \citenamefont {Campbell}, \citenamefont {DeMille}, \citenamefont {Doyle},
  \citenamefont {Gabrielse}, \citenamefont {Gurevich}, \citenamefont {Hess},
  \citenamefont {Hutzler}, \citenamefont {Kirilov}, \citenamefont {Kozyryev},
  \citenamefont {O'Leary}, \citenamefont {Panda}, \citenamefont {Parsons},
  \citenamefont {Petrik}, \citenamefont {Spaun}, \citenamefont {Vutha},\ and\
  \citenamefont {West}}]{Baron2013}%
  \BibitemOpen
  \bibfield  {author} {\bibinfo {author} {\bibfnamefont {J.}~\bibnamefont
  {Baron}}, \bibinfo {author} {\bibfnamefont {W.~C.}\ \bibnamefont {Campbell}},
  \bibinfo {author} {\bibfnamefont {D.}~\bibnamefont {DeMille}}, \bibinfo
  {author} {\bibfnamefont {J.~M.}\ \bibnamefont {Doyle}}, \bibinfo {author}
  {\bibfnamefont {G.}~\bibnamefont {Gabrielse}}, \bibinfo {author}
  {\bibfnamefont {Y.~V.}\ \bibnamefont {Gurevich}}, \bibinfo {author}
  {\bibfnamefont {P.~W.}\ \bibnamefont {Hess}}, \bibinfo {author}
  {\bibfnamefont {N.~R.}\ \bibnamefont {Hutzler}}, \bibinfo {author}
  {\bibfnamefont {E.}~\bibnamefont {Kirilov}}, \bibinfo {author} {\bibfnamefont
  {I.}~\bibnamefont {Kozyryev}}, \bibinfo {author} {\bibfnamefont {B.~R.}\
  \bibnamefont {O'Leary}}, \bibinfo {author} {\bibfnamefont {C.~D.}\
  \bibnamefont {Panda}}, \bibinfo {author} {\bibfnamefont {M.~F.}\ \bibnamefont
  {Parsons}}, \bibinfo {author} {\bibfnamefont {E.~S.}\ \bibnamefont {Petrik}},
  \bibinfo {author} {\bibfnamefont {B.}~\bibnamefont {Spaun}}, \bibinfo
  {author} {\bibfnamefont {A.~C.}\ \bibnamefont {Vutha}}, \ and\ \bibinfo
  {author} {\bibfnamefont {A.~D.}\ \bibnamefont {West}},\ }\href@noop {} {}
  (\bibinfo {year} {2013}),\ \Eprint {http://arxiv.org/abs/1310.7534}
  {arXiv:1310.7534 [physics.atom-ph]} \BibitemShut {NoStop}%
\bibitem [{\citenamefont {Hudson}\ \emph {et~al.}(2006)\citenamefont {Hudson},
  \citenamefont {Ticknor}, \citenamefont {Sawyer}, \citenamefont {Taatjes},
  \citenamefont {Lewandowski}, \citenamefont {Bochinski}, \citenamefont
  {Bohn},\ and\ \citenamefont {Ye}}]{Hudson:2006a}%
  \BibitemOpen
  \bibfield  {author} {\bibinfo {author} {\bibfnamefont {E.~R.}\ \bibnamefont
  {Hudson}}, \bibinfo {author} {\bibfnamefont {C.}~\bibnamefont {Ticknor}},
  \bibinfo {author} {\bibfnamefont {B.~C.}\ \bibnamefont {Sawyer}}, \bibinfo
  {author} {\bibfnamefont {C.~A.}\ \bibnamefont {Taatjes}}, \bibinfo {author}
  {\bibfnamefont {H.~J.}\ \bibnamefont {Lewandowski}}, \bibinfo {author}
  {\bibfnamefont {J.~R.}\ \bibnamefont {Bochinski}}, \bibinfo {author}
  {\bibfnamefont {J.~L.}\ \bibnamefont {Bohn}}, \ and\ \bibinfo {author}
  {\bibfnamefont {J.}~\bibnamefont {Ye}},\ }\href@noop {} {\bibfield  {journal}
  {\bibinfo  {journal} {Phys. Rev. A}\ }\textbf {\bibinfo {volume} {73}},\
  \bibinfo {pages} {063404} (\bibinfo {year} {2006})}\BibitemShut {NoStop}%
\bibitem [{\citenamefont {Krems}(2008)}]{Krems:2008}%
  \BibitemOpen
  \bibfield  {author} {\bibinfo {author} {\bibfnamefont {R.~V.}\ \bibnamefont
  {Krems}},\ }\href@noop {} {\bibfield  {journal} {\bibinfo  {journal} {Phys.
  Chem. Chem. Phys.}\ }\textbf {\bibinfo {volume} {10}},\ \bibinfo {pages}
  {4079} (\bibinfo {year} {2008})}\BibitemShut {NoStop}%
\bibitem [{\citenamefont {Bethlem}\ and\ \citenamefont
  {Meijer}(2003)}]{Bethlem:IRPC:2003}%
  \BibitemOpen
  \bibfield  {author} {\bibinfo {author} {\bibfnamefont {H.~L.}\ \bibnamefont
  {Bethlem}}\ and\ \bibinfo {author} {\bibfnamefont {G.}~\bibnamefont
  {Meijer}},\ }\href@noop {} {\bibfield  {journal} {\bibinfo  {journal} {Int.
  Rev. Phys. Chem.}\ }\textbf {\bibinfo {volume} {22}},\ \bibinfo {pages} {73}
  (\bibinfo {year} {2003})}\BibitemShut {NoStop}%
\bibitem [{\citenamefont {Sage}\ \emph {et~al.}(2005)\citenamefont {Sage},
  \citenamefont {Sainis}, \citenamefont {Bergeman},\ and\ \citenamefont
  {DeMille}}]{Sage:2005}%
  \BibitemOpen
  \bibfield  {author} {\bibinfo {author} {\bibfnamefont {J.~M.}\ \bibnamefont
  {Sage}}, \bibinfo {author} {\bibfnamefont {S.}~\bibnamefont {Sainis}},
  \bibinfo {author} {\bibfnamefont {T.}~\bibnamefont {Bergeman}}, \ and\
  \bibinfo {author} {\bibfnamefont {D.}~\bibnamefont {DeMille}},\ }\href@noop
  {} {\bibfield  {journal} {\bibinfo  {journal} {Phys. Rev. Lett.}\ }\textbf
  {\bibinfo {volume} {94}},\ \bibinfo {pages} {203001} (\bibinfo {year}
  {2005})}\BibitemShut {NoStop}%
\bibitem [{\citenamefont {Aikawa}\ \emph {et~al.}(2010)\citenamefont {Aikawa},
  \citenamefont {Akamatsu}, \citenamefont {Hayashi}, \citenamefont {Oasa},
  \citenamefont {Kobayashi}, \citenamefont {Naidon}, \citenamefont {Kishimoto},
  \citenamefont {Ueda},\ and\ \citenamefont {Inouye}}]{Aikawa:2010}%
  \BibitemOpen
  \bibfield  {author} {\bibinfo {author} {\bibfnamefont {K.}~\bibnamefont
  {Aikawa}}, \bibinfo {author} {\bibfnamefont {D.}~\bibnamefont {Akamatsu}},
  \bibinfo {author} {\bibfnamefont {M.}~\bibnamefont {Hayashi}}, \bibinfo
  {author} {\bibfnamefont {K.}~\bibnamefont {Oasa}}, \bibinfo {author}
  {\bibfnamefont {J.}~\bibnamefont {Kobayashi}}, \bibinfo {author}
  {\bibfnamefont {P.}~\bibnamefont {Naidon}}, \bibinfo {author} {\bibfnamefont
  {T.}~\bibnamefont {Kishimoto}}, \bibinfo {author} {\bibfnamefont
  {M.}~\bibnamefont {Ueda}}, \ and\ \bibinfo {author} {\bibfnamefont
  {S.}~\bibnamefont {Inouye}},\ }\href {\doibase
  10.1103/PhysRevLett.105.203001} {\bibfield  {journal} {\bibinfo  {journal}
  {Phys. Rev. Lett.}\ }\textbf {\bibinfo {volume} {105}},\ \bibinfo {pages}
  {203001} (\bibinfo {year} {2010})}\BibitemShut {NoStop}%
\bibitem [{\citenamefont {Barry}\ \emph {et~al.}(2012)\citenamefont {Barry},
  \citenamefont {Shuman}, \citenamefont {Norrgard},\ and\ \citenamefont
  {DeMille}}]{Barry:2012}%
  \BibitemOpen
  \bibfield  {author} {\bibinfo {author} {\bibfnamefont {J.~F.}\ \bibnamefont
  {Barry}}, \bibinfo {author} {\bibfnamefont {E.~S.}\ \bibnamefont {Shuman}},
  \bibinfo {author} {\bibfnamefont {E.~B.}\ \bibnamefont {Norrgard}}, \ and\
  \bibinfo {author} {\bibfnamefont {D.}~\bibnamefont {DeMille}},\ }\href@noop
  {} {\bibfield  {journal} {\bibinfo  {journal} {Phys. Rev. Lett.}\ }\textbf
  {\bibinfo {volume} {108}},\ \bibinfo {pages} {103002} (\bibinfo {year}
  {2012})}\BibitemShut {NoStop}%
\bibitem [{\citenamefont {Damski}\ \emph {et~al.}(2003)\citenamefont {Damski},
  \citenamefont {Santos}, \citenamefont {Tiemann}, \citenamefont {Lewenstein},
  \citenamefont {Kotochigova}, \citenamefont {Julienne},\ and\ \citenamefont
  {Zoller}}]{Damski:2003}%
  \BibitemOpen
  \bibfield  {author} {\bibinfo {author} {\bibfnamefont {B.}~\bibnamefont
  {Damski}}, \bibinfo {author} {\bibfnamefont {L.}~\bibnamefont {Santos}},
  \bibinfo {author} {\bibfnamefont {E.}~\bibnamefont {Tiemann}}, \bibinfo
  {author} {\bibfnamefont {M.}~\bibnamefont {Lewenstein}}, \bibinfo {author}
  {\bibfnamefont {S.}~\bibnamefont {Kotochigova}}, \bibinfo {author}
  {\bibfnamefont {P.}~\bibnamefont {Julienne}}, \ and\ \bibinfo {author}
  {\bibfnamefont {P.}~\bibnamefont {Zoller}},\ }\href@noop {} {\bibfield
  {journal} {\bibinfo  {journal} {Phys. Rev. Lett.}\ }\textbf {\bibinfo
  {volume} {90}},\ \bibinfo {pages} {110401} (\bibinfo {year}
  {2003})}\BibitemShut {NoStop}%
\bibitem [{\citenamefont {Chin}\ \emph {et~al.}(2010)\citenamefont {Chin},
  \citenamefont {Grimm}, \citenamefont {Tiesinga},\ and\ \citenamefont
  {Julienne}}]{Chin:RMP:2010}%
  \BibitemOpen
  \bibfield  {author} {\bibinfo {author} {\bibfnamefont {C.}~\bibnamefont
  {Chin}}, \bibinfo {author} {\bibfnamefont {R.}~\bibnamefont {Grimm}},
  \bibinfo {author} {\bibfnamefont {E.}~\bibnamefont {Tiesinga}}, \ and\
  \bibinfo {author} {\bibfnamefont {P.~S.}\ \bibnamefont {Julienne}},\
  }\href@noop {} {\bibfield  {journal} {\bibinfo  {journal} {Rev. Mod. Phys.}\
  }\textbf {\bibinfo {volume} {82}},\ \bibinfo {pages} {1225} (\bibinfo {year}
  {2010})}\BibitemShut {NoStop}%
\bibitem [{\citenamefont {Bergmann}\ \emph {et~al.}(1998)\citenamefont
  {Bergmann}, \citenamefont {Theuer},\ and\ \citenamefont
  {Shore}}]{Bergmann:1998}%
  \BibitemOpen
  \bibfield  {author} {\bibinfo {author} {\bibfnamefont {K.}~\bibnamefont
  {Bergmann}}, \bibinfo {author} {\bibfnamefont {H.}~\bibnamefont {Theuer}}, \
  and\ \bibinfo {author} {\bibfnamefont {B.~W.}\ \bibnamefont {Shore}},\
  }\href@noop {} {\bibfield  {journal} {\bibinfo  {journal} {Rev. Mod. Phys.}\
  }\textbf {\bibinfo {volume} {70}},\ \bibinfo {pages} {1003} (\bibinfo {year}
  {1998})}\BibitemShut {NoStop}%
\bibitem [{\citenamefont {Ni}\ \emph {et~al.}(2008)\citenamefont {Ni},
  \citenamefont {Ospelkaus}, \citenamefont {{de Miranda}}, \citenamefont
  {Pe'er}, \citenamefont {Neyenhuis}, \citenamefont {Zirbel}, \citenamefont
  {Kotochigova}, \citenamefont {Julienne}, \citenamefont {Jin},\ and\
  \citenamefont {Ye}}]{Ni:KRb:2008}%
  \BibitemOpen
  \bibfield  {author} {\bibinfo {author} {\bibfnamefont {K.-K.}\ \bibnamefont
  {Ni}}, \bibinfo {author} {\bibfnamefont {S.}~\bibnamefont {Ospelkaus}},
  \bibinfo {author} {\bibfnamefont {M.~H.~G.}\ \bibnamefont {{de Miranda}}},
  \bibinfo {author} {\bibfnamefont {A.}~\bibnamefont {Pe'er}}, \bibinfo
  {author} {\bibfnamefont {B.}~\bibnamefont {Neyenhuis}}, \bibinfo {author}
  {\bibfnamefont {J.~J.}\ \bibnamefont {Zirbel}}, \bibinfo {author}
  {\bibfnamefont {S.}~\bibnamefont {Kotochigova}}, \bibinfo {author}
  {\bibfnamefont {P.~S.}\ \bibnamefont {Julienne}}, \bibinfo {author}
  {\bibfnamefont {D.~S.}\ \bibnamefont {Jin}}, \ and\ \bibinfo {author}
  {\bibfnamefont {J.}~\bibnamefont {Ye}},\ }\href@noop {} {\bibfield  {journal}
  {\bibinfo  {journal} {Science}\ }\textbf {\bibinfo {volume} {322}},\ \bibinfo
  {pages} {231} (\bibinfo {year} {2008})}\BibitemShut {NoStop}%
\bibitem [{\citenamefont {Lang}\ \emph {et~al.}(2008)\citenamefont {Lang},
  \citenamefont {Winkler}, \citenamefont {Strauss}, \citenamefont {Grimm},\
  and\ \citenamefont {Hecker~Denschlag}}]{Lang:ground:2008}%
  \BibitemOpen
  \bibfield  {author} {\bibinfo {author} {\bibfnamefont {F.}~\bibnamefont
  {Lang}}, \bibinfo {author} {\bibfnamefont {K.}~\bibnamefont {Winkler}},
  \bibinfo {author} {\bibfnamefont {C.}~\bibnamefont {Strauss}}, \bibinfo
  {author} {\bibfnamefont {R.}~\bibnamefont {Grimm}}, \ and\ \bibinfo {author}
  {\bibfnamefont {J.}~\bibnamefont {Hecker~Denschlag}},\ }\href@noop {}
  {\bibfield  {journal} {\bibinfo  {journal} {Phys. Rev. Lett.}\ }\textbf
  {\bibinfo {volume} {101}},\ \bibinfo {pages} {133005} (\bibinfo {year}
  {2008})}\BibitemShut {NoStop}%
\bibitem [{\citenamefont {Danzl}\ \emph {et~al.}(2010)\citenamefont {Danzl},
  \citenamefont {Mark}, \citenamefont {Haller}, \citenamefont {Gustavsson},
  \citenamefont {Hart}, \citenamefont {Aldegunde}, \citenamefont {Hutson},\
  and\ \citenamefont {N\"agerl}}]{Danzl:ground:2010}%
  \BibitemOpen
  \bibfield  {author} {\bibinfo {author} {\bibfnamefont {J.~G.}\ \bibnamefont
  {Danzl}}, \bibinfo {author} {\bibfnamefont {M.~J.}\ \bibnamefont {Mark}},
  \bibinfo {author} {\bibfnamefont {E.}~\bibnamefont {Haller}}, \bibinfo
  {author} {\bibfnamefont {M.}~\bibnamefont {Gustavsson}}, \bibinfo {author}
  {\bibfnamefont {R.}~\bibnamefont {Hart}}, \bibinfo {author} {\bibfnamefont
  {J.}~\bibnamefont {Aldegunde}}, \bibinfo {author} {\bibfnamefont {J.~M.}\
  \bibnamefont {Hutson}}, \ and\ \bibinfo {author} {\bibfnamefont {H.-C.}\
  \bibnamefont {N\"agerl}},\ }\href {\doibase doi:10.1038/nphys1533} {\bibfield
   {journal} {\bibinfo  {journal} {Nature Phys.}\ }\textbf {\bibinfo {volume}
  {6}},\ \bibinfo {pages} {265} (\bibinfo {year} {2010})}\BibitemShut {NoStop}%
\bibitem [{\citenamefont {Ospelkaus}\ \emph {et~al.}(2010)\citenamefont
  {Ospelkaus}, \citenamefont {Ni}, \citenamefont {Wang}, \citenamefont {{de
  Miranda}}, \citenamefont {Neyenhuis}, \citenamefont {Qu\'{e}m\'{e}ner},
  \citenamefont {Julienne}, \citenamefont {Bohn}, \citenamefont {Jin},\ and\
  \citenamefont {Ye}}]{Ospelkaus:react:2010}%
  \BibitemOpen
  \bibfield  {author} {\bibinfo {author} {\bibfnamefont {S.}~\bibnamefont
  {Ospelkaus}}, \bibinfo {author} {\bibfnamefont {K.-K.}\ \bibnamefont {Ni}},
  \bibinfo {author} {\bibfnamefont {D.}~\bibnamefont {Wang}}, \bibinfo {author}
  {\bibfnamefont {M.~H.~G.}\ \bibnamefont {{de Miranda}}}, \bibinfo {author}
  {\bibfnamefont {B.}~\bibnamefont {Neyenhuis}}, \bibinfo {author}
  {\bibfnamefont {G.}~\bibnamefont {Qu\'{e}m\'{e}ner}}, \bibinfo {author}
  {\bibfnamefont {P.~S.}\ \bibnamefont {Julienne}}, \bibinfo {author}
  {\bibfnamefont {J.~L.}\ \bibnamefont {Bohn}}, \bibinfo {author}
  {\bibfnamefont {D.~S.}\ \bibnamefont {Jin}}, \ and\ \bibinfo {author}
  {\bibfnamefont {J.}~\bibnamefont {Ye}},\ }\href@noop {} {\bibfield  {journal}
  {\bibinfo  {journal} {Science}\ }\textbf {\bibinfo {volume} {327}},\ \bibinfo
  {pages} {853} (\bibinfo {year} {2010})}\BibitemShut {NoStop}%
\bibitem [{\citenamefont {\.Zuchowski}\ and\ \citenamefont
  {Hutson}(2010)}]{Zuchowski:trimers:2010}%
  \BibitemOpen
  \bibfield  {author} {\bibinfo {author} {\bibfnamefont {P.~S.}\ \bibnamefont
  {\.Zuchowski}}\ and\ \bibinfo {author} {\bibfnamefont {J.~M.}\ \bibnamefont
  {Hutson}},\ }\href@noop {} {\bibfield  {journal} {\bibinfo  {journal} {Phys.
  Rev. A}\ }\textbf {\bibinfo {volume} {81}},\ \bibinfo {pages} {060703(R)}
  (\bibinfo {year} {2010})}\BibitemShut {NoStop}%
\bibitem [{\citenamefont {Pilch}\ \emph {et~al.}(2009)\citenamefont {Pilch},
  \citenamefont {Lange}, \citenamefont {Prantner}, \citenamefont {Kerner},
  \citenamefont {Ferlaino}, \citenamefont {N\"agerl},\ and\ \citenamefont
  {Grimm}}]{Pilch:2009}%
  \BibitemOpen
  \bibfield  {author} {\bibinfo {author} {\bibfnamefont {K.}~\bibnamefont
  {Pilch}}, \bibinfo {author} {\bibfnamefont {A.~D.}\ \bibnamefont {Lange}},
  \bibinfo {author} {\bibfnamefont {A.}~\bibnamefont {Prantner}}, \bibinfo
  {author} {\bibfnamefont {G.}~\bibnamefont {Kerner}}, \bibinfo {author}
  {\bibfnamefont {F.}~\bibnamefont {Ferlaino}}, \bibinfo {author}
  {\bibfnamefont {H.-C.}\ \bibnamefont {N\"agerl}}, \ and\ \bibinfo {author}
  {\bibfnamefont {R.}~\bibnamefont {Grimm}},\ }\href@noop {} {\bibfield
  {journal} {\bibinfo  {journal} {Phys. Rev. A}\ }\textbf {\bibinfo {volume}
  {79}},\ \bibinfo {pages} {042718} (\bibinfo {year} {2009})}\BibitemShut
  {NoStop}%
\bibitem [{\citenamefont {Lercher}\ \emph {et~al.}(2011)\citenamefont
  {Lercher}, \citenamefont {Takekoshi}, \citenamefont {Debatin}, \citenamefont
  {Schuster}, \citenamefont {Rameshan}, \citenamefont {Ferlaino}, \citenamefont
  {Grimm},\ and\ \citenamefont {N\"agerl}}]{Lercher:2011}%
  \BibitemOpen
  \bibfield  {author} {\bibinfo {author} {\bibfnamefont {A.~D.}\ \bibnamefont
  {Lercher}}, \bibinfo {author} {\bibfnamefont {T.}~\bibnamefont {Takekoshi}},
  \bibinfo {author} {\bibfnamefont {M.}~\bibnamefont {Debatin}}, \bibinfo
  {author} {\bibfnamefont {B.}~\bibnamefont {Schuster}}, \bibinfo {author}
  {\bibfnamefont {R.}~\bibnamefont {Rameshan}}, \bibinfo {author}
  {\bibfnamefont {F.}~\bibnamefont {Ferlaino}}, \bibinfo {author}
  {\bibfnamefont {R.}~\bibnamefont {Grimm}}, \ and\ \bibinfo {author}
  {\bibfnamefont {H.-C.}\ \bibnamefont {N\"agerl}},\ }\href {\doibase
  10.1140/epjd/e2011-20015-6} {\bibfield  {journal} {\bibinfo  {journal} {Eur.
  Phys. J. D}\ }\textbf {\bibinfo {volume} {65}},\ \bibinfo {pages} {3}
  (\bibinfo {year} {2011})}\BibitemShut {NoStop}%
\bibitem [{\citenamefont {McCarron}\ \emph {et~al.}(2011)\citenamefont
  {McCarron}, \citenamefont {Cho}, \citenamefont {Jenkin}, \citenamefont
  {K\"oppinger},\ and\ \citenamefont {Cornish}}]{McCarron:2011}%
  \BibitemOpen
  \bibfield  {author} {\bibinfo {author} {\bibfnamefont {D.~J.}\ \bibnamefont
  {McCarron}}, \bibinfo {author} {\bibfnamefont {H.~W.}\ \bibnamefont {Cho}},
  \bibinfo {author} {\bibfnamefont {D.~L.}\ \bibnamefont {Jenkin}}, \bibinfo
  {author} {\bibfnamefont {M.~P.}\ \bibnamefont {K\"oppinger}}, \ and\ \bibinfo
  {author} {\bibfnamefont {S.~L.}\ \bibnamefont {Cornish}},\ }\href@noop {}
  {\bibfield  {journal} {\bibinfo  {journal} {Phys. Rev. A}\ }\textbf {\bibinfo
  {volume} {84}},\ \bibinfo {pages} {011603} (\bibinfo {year}
  {2011})}\BibitemShut {NoStop}%
\bibitem [{\citenamefont {Pattinson}\ \emph {et~al.}(2013)\citenamefont
  {Pattinson}, \citenamefont {Billam}, \citenamefont {Gardiner}, \citenamefont
  {McCarron}, \citenamefont {Cho}, \citenamefont {Cornish}, \citenamefont
  {Parker},\ and\ \citenamefont {Proukakis}}]{Pattinson2013}%
  \BibitemOpen
  \bibfield  {author} {\bibinfo {author} {\bibfnamefont {R.~W.}\ \bibnamefont
  {Pattinson}}, \bibinfo {author} {\bibfnamefont {T.~P.}\ \bibnamefont
  {Billam}}, \bibinfo {author} {\bibfnamefont {S.~A.}\ \bibnamefont
  {Gardiner}}, \bibinfo {author} {\bibfnamefont {D.~J.}\ \bibnamefont
  {McCarron}}, \bibinfo {author} {\bibfnamefont {H.~W.}\ \bibnamefont {Cho}},
  \bibinfo {author} {\bibfnamefont {S.~L.}\ \bibnamefont {Cornish}}, \bibinfo
  {author} {\bibfnamefont {N.~G.}\ \bibnamefont {Parker}}, \ and\ \bibinfo
  {author} {\bibfnamefont {N.~P.}\ \bibnamefont {Proukakis}},\ }\href {\doibase
  10.1103/PhysRevA.87.013625} {\bibfield  {journal} {\bibinfo  {journal} {Phys.
  Rev. A}\ }\textbf {\bibinfo {volume} {87}},\ \bibinfo {pages} {013625}
  (\bibinfo {year} {2013})}\BibitemShut {NoStop}%
\bibitem [{\citenamefont {Takekoshi}\ \emph {et~al.}(2012)\citenamefont
  {Takekoshi}, \citenamefont {Debatin}, \citenamefont {Rameshan}, \citenamefont
  {Ferlaino}, \citenamefont {Grimm}, \citenamefont {N\"agerl}, \citenamefont
  {{Le Sueur}}, \citenamefont {Hutson}, \citenamefont {Julienne}, \citenamefont
  {Kotochigova},\ and\ \citenamefont {Tiemann}}]{Takekoshi:RbCs:2012}%
  \BibitemOpen
  \bibfield  {author} {\bibinfo {author} {\bibfnamefont {T.}~\bibnamefont
  {Takekoshi}}, \bibinfo {author} {\bibfnamefont {M.}~\bibnamefont {Debatin}},
  \bibinfo {author} {\bibfnamefont {R.}~\bibnamefont {Rameshan}}, \bibinfo
  {author} {\bibfnamefont {F.}~\bibnamefont {Ferlaino}}, \bibinfo {author}
  {\bibfnamefont {R.}~\bibnamefont {Grimm}}, \bibinfo {author} {\bibfnamefont
  {H.-C.}\ \bibnamefont {N\"agerl}}, \bibinfo {author} {\bibfnamefont {C.~R.}\
  \bibnamefont {{Le Sueur}}}, \bibinfo {author} {\bibfnamefont {J.~M.}\
  \bibnamefont {Hutson}}, \bibinfo {author} {\bibfnamefont {P.~S.}\
  \bibnamefont {Julienne}}, \bibinfo {author} {\bibfnamefont {S.}~\bibnamefont
  {Kotochigova}}, \ and\ \bibinfo {author} {\bibfnamefont {E.}~\bibnamefont
  {Tiemann}},\ }\href@noop {} {\bibfield  {journal} {\bibinfo  {journal} {Phys.
  Rev. A}\ }\textbf {\bibinfo {volume} {85}},\ \bibinfo {pages} {032506}
  (\bibinfo {year} {2012})}\BibitemShut {NoStop}%
\bibitem [{\citenamefont {Cho}\ \emph {et~al.}(2013)\citenamefont {Cho},
  \citenamefont {McCarron}, \citenamefont {K\"oppinger}, \citenamefont
  {Jenkin}, \citenamefont {Butler}, \citenamefont {Julienne}, \citenamefont
  {Blackley}, \citenamefont {Le~Sueur}, \citenamefont {Hutson},\ and\
  \citenamefont {Cornish}}]{Cho2013}%
  \BibitemOpen
  \bibfield  {author} {\bibinfo {author} {\bibfnamefont {H.-W.}\ \bibnamefont
  {Cho}}, \bibinfo {author} {\bibfnamefont {D.~J.}\ \bibnamefont {McCarron}},
  \bibinfo {author} {\bibfnamefont {M.~P.}\ \bibnamefont {K\"oppinger}},
  \bibinfo {author} {\bibfnamefont {D.~L.}\ \bibnamefont {Jenkin}}, \bibinfo
  {author} {\bibfnamefont {K.~L.}\ \bibnamefont {Butler}}, \bibinfo {author}
  {\bibfnamefont {P.~S.}\ \bibnamefont {Julienne}}, \bibinfo {author}
  {\bibfnamefont {C.~L.}\ \bibnamefont {Blackley}}, \bibinfo {author}
  {\bibfnamefont {C.~R.}\ \bibnamefont {Le~Sueur}}, \bibinfo {author}
  {\bibfnamefont {J.~M.}\ \bibnamefont {Hutson}}, \ and\ \bibinfo {author}
  {\bibfnamefont {S.~L.}\ \bibnamefont {Cornish}},\ }\href {\doibase
  10.1103/PhysRevA.87.010703} {\bibfield  {journal} {\bibinfo  {journal} {Phys.
  Rev. A}\ }\textbf {\bibinfo {volume} {87}},\ \bibinfo {pages} {010703}
  (\bibinfo {year} {2013})}\BibitemShut {NoStop}%
\bibitem [{\citenamefont {Matthews}\ \emph {et~al.}(1999)\citenamefont
  {Matthews}, \citenamefont {Anderson}, \citenamefont {Haljan}, \citenamefont
  {Hall}, \citenamefont {Wieman},\ and\ \citenamefont
  {Cornell}}]{Matthews:1999}%
  \BibitemOpen
  \bibfield  {author} {\bibinfo {author} {\bibfnamefont {M.~R.}\ \bibnamefont
  {Matthews}}, \bibinfo {author} {\bibfnamefont {B.~P.}\ \bibnamefont
  {Anderson}}, \bibinfo {author} {\bibfnamefont {P.~C.}\ \bibnamefont
  {Haljan}}, \bibinfo {author} {\bibfnamefont {D.~S.}\ \bibnamefont {Hall}},
  \bibinfo {author} {\bibfnamefont {C.~E.}\ \bibnamefont {Wieman}}, \ and\
  \bibinfo {author} {\bibfnamefont {E.~A.}\ \bibnamefont {Cornell}},\
  }\href@noop {} {\bibfield  {journal} {\bibinfo  {journal} {Phys. Rev. Lett.}\
  }\textbf {\bibinfo {volume} {83}},\ \bibinfo {pages} {2498} (\bibinfo {year}
  {1999})}\BibitemShut {NoStop}%
\bibitem [{\citenamefont {Papp}\ \emph {et~al.}(2008)\citenamefont {Papp},
  \citenamefont {Pino},\ and\ \citenamefont {Wieman}}]{Papp2008}%
  \BibitemOpen
  \bibfield  {author} {\bibinfo {author} {\bibfnamefont {S.~B.}\ \bibnamefont
  {Papp}}, \bibinfo {author} {\bibfnamefont {J.~M.}\ \bibnamefont {Pino}}, \
  and\ \bibinfo {author} {\bibfnamefont {C.~E.}\ \bibnamefont {Wieman}},\
  }\href {\doibase 10.1103/PhysRevLett.101.040402} {\bibfield  {journal}
  {\bibinfo  {journal} {Phys. Rev. Lett.}\ }\textbf {\bibinfo {volume} {101}},\
  \bibinfo {pages} {040402} (\bibinfo {year} {2008})}\BibitemShut {NoStop}%
\bibitem [{\citenamefont {Becker}\ \emph {et~al.}(2008)\citenamefont {Becker},
  \citenamefont {Stellmer}, \citenamefont {Soltan-Panahi}, \citenamefont
  {Dorscher}, \citenamefont {Baumert}, \citenamefont {Richter}, \citenamefont
  {Kronjager}, \citenamefont {Bongs},\ and\ \citenamefont
  {Sengstock}}]{Becker2008}%
  \BibitemOpen
  \bibfield  {author} {\bibinfo {author} {\bibfnamefont {C.}~\bibnamefont
  {Becker}}, \bibinfo {author} {\bibfnamefont {S.}~\bibnamefont {Stellmer}},
  \bibinfo {author} {\bibfnamefont {P.}~\bibnamefont {Soltan-Panahi}}, \bibinfo
  {author} {\bibfnamefont {S.}~\bibnamefont {Dorscher}}, \bibinfo {author}
  {\bibfnamefont {M.}~\bibnamefont {Baumert}}, \bibinfo {author} {\bibfnamefont
  {E.~M.}\ \bibnamefont {Richter}}, \bibinfo {author} {\bibfnamefont
  {J.}~\bibnamefont {Kronjager}}, \bibinfo {author} {\bibfnamefont
  {K.}~\bibnamefont {Bongs}}, \ and\ \bibinfo {author} {\bibfnamefont
  {K.}~\bibnamefont {Sengstock}},\ }\href {\doibase 10.1038/nphys962}
  {\bibfield  {journal} {\bibinfo  {journal} {Nature Phys.}\ }\textbf {\bibinfo
  {volume} {4}},\ \bibinfo {pages} {469} (\bibinfo {year} {2008})}\BibitemShut
  {NoStop}%
\bibitem [{\citenamefont {Leo}\ \emph {et~al.}(2000)\citenamefont {Leo},
  \citenamefont {Williams},\ and\ \citenamefont {Julienne}}]{Leo:2000}%
  \BibitemOpen
  \bibfield  {author} {\bibinfo {author} {\bibfnamefont {P.~J.}\ \bibnamefont
  {Leo}}, \bibinfo {author} {\bibfnamefont {C.~J.}\ \bibnamefont {Williams}}, \
  and\ \bibinfo {author} {\bibfnamefont {P.~S.}\ \bibnamefont {Julienne}},\
  }\href@noop {} {\bibfield  {journal} {\bibinfo  {journal} {Phys. Rev. Lett.}\
  }\textbf {\bibinfo {volume} {85}},\ \bibinfo {pages} {2721} (\bibinfo {year}
  {2000})}\BibitemShut {NoStop}%
\bibitem [{\citenamefont {Chin}\ \emph {et~al.}(2000)\citenamefont {Chin},
  \citenamefont {Vuleti\'{c}}, \citenamefont {Kerman},\ and\ \citenamefont
  {Chu}}]{Chin2000}%
  \BibitemOpen
  \bibfield  {author} {\bibinfo {author} {\bibfnamefont {C.}~\bibnamefont
  {Chin}}, \bibinfo {author} {\bibfnamefont {V.}~\bibnamefont {Vuleti\'{c}}},
  \bibinfo {author} {\bibfnamefont {A.~J.}\ \bibnamefont {Kerman}}, \ and\
  \bibinfo {author} {\bibfnamefont {S.}~\bibnamefont {Chu}},\ }\href@noop {}
  {\bibfield  {journal} {\bibinfo  {journal} {Phys. Rev. Lett.}\ }\textbf
  {\bibinfo {volume} {85}},\ \bibinfo {pages} {2717} (\bibinfo {year}
  {2000})}\BibitemShut {NoStop}%
\bibitem [{\citenamefont {Chin}\ \emph {et~al.}(2004)\citenamefont {Chin},
  \citenamefont {Vuleti\'c}, \citenamefont {Kerman}, \citenamefont {Chu},
  \citenamefont {Tiesinga}, \citenamefont {Leo},\ and\ \citenamefont
  {Williams}}]{Chin:cs2-fesh:2004}%
  \BibitemOpen
  \bibfield  {author} {\bibinfo {author} {\bibfnamefont {C.}~\bibnamefont
  {Chin}}, \bibinfo {author} {\bibfnamefont {V.}~\bibnamefont {Vuleti\'c}},
  \bibinfo {author} {\bibfnamefont {A.~J.}\ \bibnamefont {Kerman}}, \bibinfo
  {author} {\bibfnamefont {S.}~\bibnamefont {Chu}}, \bibinfo {author}
  {\bibfnamefont {E.}~\bibnamefont {Tiesinga}}, \bibinfo {author}
  {\bibfnamefont {P.~J.}\ \bibnamefont {Leo}}, \ and\ \bibinfo {author}
  {\bibfnamefont {C.~J.}\ \bibnamefont {Williams}},\ }\href@noop {} {\bibfield
  {journal} {\bibinfo  {journal} {Phys. Rev. A}\ }\textbf {\bibinfo {volume}
  {70}},\ \bibinfo {pages} {032701} (\bibinfo {year} {2004})}\BibitemShut
  {NoStop}%
\bibitem [{\citenamefont {Berninger}\ \emph {et~al.}(2013)\citenamefont
  {Berninger}, \citenamefont {Zenesini}, \citenamefont {Huang}, \citenamefont
  {Harm}, \citenamefont {N\"agerl}, \citenamefont {Ferlaino}, \citenamefont
  {Grimm}, \citenamefont {Julienne},\ and\ \citenamefont
  {Hutson}}]{Berninger2013}%
  \BibitemOpen
  \bibfield  {author} {\bibinfo {author} {\bibfnamefont {M.}~\bibnamefont
  {Berninger}}, \bibinfo {author} {\bibfnamefont {A.}~\bibnamefont {Zenesini}},
  \bibinfo {author} {\bibfnamefont {B.}~\bibnamefont {Huang}}, \bibinfo
  {author} {\bibfnamefont {W.}~\bibnamefont {Harm}}, \bibinfo {author}
  {\bibfnamefont {H.-C.}\ \bibnamefont {N\"agerl}}, \bibinfo {author}
  {\bibfnamefont {F.}~\bibnamefont {Ferlaino}}, \bibinfo {author}
  {\bibfnamefont {R.}~\bibnamefont {Grimm}}, \bibinfo {author} {\bibfnamefont
  {P.~S.}\ \bibnamefont {Julienne}}, \ and\ \bibinfo {author} {\bibfnamefont
  {J.~M.}\ \bibnamefont {Hutson}},\ }\href {\doibase
  10.1103/PhysRevA.87.032517} {\bibfield  {journal} {\bibinfo  {journal} {Phys.
  Rev. A}\ }\textbf {\bibinfo {volume} {87}},\ \bibinfo {pages} {032517}
  (\bibinfo {year} {2013})}\BibitemShut {NoStop}%
\bibitem [{\citenamefont {Weber}\ \emph {et~al.}(2003)\citenamefont {Weber},
  \citenamefont {Herbig}, \citenamefont {Mark}, \citenamefont {N\"{a}gerl},\
  and\ \citenamefont {Grimm}}]{Weber:CsBEC:2003}%
  \BibitemOpen
  \bibfield  {author} {\bibinfo {author} {\bibfnamefont {T.}~\bibnamefont
  {Weber}}, \bibinfo {author} {\bibfnamefont {J.}~\bibnamefont {Herbig}},
  \bibinfo {author} {\bibfnamefont {M.}~\bibnamefont {Mark}}, \bibinfo {author}
  {\bibfnamefont {H.~C.}\ \bibnamefont {N\"{a}gerl}}, \ and\ \bibinfo {author}
  {\bibfnamefont {R.}~\bibnamefont {Grimm}},\ }\href@noop {} {\bibfield
  {journal} {\bibinfo  {journal} {Science}\ }\textbf {\bibinfo {volume}
  {299}},\ \bibinfo {pages} {232} (\bibinfo {year} {2003})}\BibitemShut
  {NoStop}%
\bibitem [{\citenamefont {Hung}\ \emph {et~al.}(2008)\citenamefont {Hung},
  \citenamefont {Zhang}, \citenamefont {Gemelke},\ and\ \citenamefont
  {Chin}}]{Hung2008}%
  \BibitemOpen
  \bibfield  {author} {\bibinfo {author} {\bibfnamefont {C.-L.}\ \bibnamefont
  {Hung}}, \bibinfo {author} {\bibfnamefont {X.}~\bibnamefont {Zhang}},
  \bibinfo {author} {\bibfnamefont {N.}~\bibnamefont {Gemelke}}, \ and\
  \bibinfo {author} {\bibfnamefont {C.}~\bibnamefont {Chin}},\ }\href {\doibase
  10.1103/PhysRevA.78.011604} {\bibfield  {journal} {\bibinfo  {journal} {Phys.
  Rev. A}\ }\textbf {\bibinfo {volume} {78}},\ \bibinfo {pages} {011604}
  (\bibinfo {year} {2008})}\BibitemShut {NoStop}%
\bibitem [{\citenamefont {Hutson}\ and\ \citenamefont
  {Green}(1994)}]{molscat:v14}%
  \BibitemOpen
  \bibfield  {author} {\bibinfo {author} {\bibfnamefont {J.~M.}\ \bibnamefont
  {Hutson}}\ and\ \bibinfo {author} {\bibfnamefont {S.}~\bibnamefont {Green}},\
  }\href@noop {} {\enquote {\bibinfo {title} {{MOLSCAT} computer program,
  version 14},}\ }\bibinfo {howpublished} {distributed by Collaborative
  Computational Project No.\ 6 of the UK Engineering and Physical Sciences
  Research Council} (\bibinfo {year} {1994})\BibitemShut {NoStop}%
\bibitem [{\citenamefont {Blackley}\ \emph {et~al.}(2013)\citenamefont
  {Blackley}, \citenamefont {{Le Sueur}}, \citenamefont {Hutson}, \citenamefont
  {McCarron}, \citenamefont {K\"oppinger}, \citenamefont {Cho}, \citenamefont
  {Jenkin},\ and\ \citenamefont {Cornish}}]{Blackley:85Rb:2013}%
  \BibitemOpen
  \bibfield  {author} {\bibinfo {author} {\bibfnamefont {C.~L.}\ \bibnamefont
  {Blackley}}, \bibinfo {author} {\bibfnamefont {C.~R.}\ \bibnamefont {{Le
  Sueur}}}, \bibinfo {author} {\bibfnamefont {J.~M.}\ \bibnamefont {Hutson}},
  \bibinfo {author} {\bibfnamefont {D.~J.}\ \bibnamefont {McCarron}}, \bibinfo
  {author} {\bibfnamefont {M.~P.}\ \bibnamefont {K\"oppinger}}, \bibinfo
  {author} {\bibfnamefont {H.-W.}\ \bibnamefont {Cho}}, \bibinfo {author}
  {\bibfnamefont {D.~L.}\ \bibnamefont {Jenkin}}, \ and\ \bibinfo {author}
  {\bibfnamefont {S.~L.}\ \bibnamefont {Cornish}},\ }\href@noop {} {\bibfield
  {journal} {\bibinfo  {journal} {Phys. Rev. A}\ }\textbf {\bibinfo {volume}
  {87}},\ \bibinfo {pages} {033611} (\bibinfo {year} {2013})}\BibitemShut
  {NoStop}%
\bibitem [{\citenamefont {Hutson}(2011)}]{Hutson:field:2011}%
  \BibitemOpen
  \bibfield  {author} {\bibinfo {author} {\bibfnamefont {J.~M.}\ \bibnamefont
  {Hutson}},\ }\href@noop {} {\enquote {\bibinfo {title} {{FIELD} computer
  program, version 1},}\ } (\bibinfo {year} {2011})\BibitemShut {NoStop}%
\bibitem [{\citenamefont {Hutson}(1993)}]{Hutson:bound:1993}%
  \BibitemOpen
  \bibfield  {author} {\bibinfo {author} {\bibfnamefont {J.~M.}\ \bibnamefont
  {Hutson}},\ }\href@noop {} {\enquote {\bibinfo {title} {{BOUND} computer
  program, version 5},}\ }\bibinfo {howpublished} {distributed by Collaborative
  Computational Project No.\ 6 of the UK Engineering and Physical Sciences
  Research Council} (\bibinfo {year} {1993})\BibitemShut {NoStop}%
\bibitem [{\citenamefont {Cho}\ \emph {et~al.}(2011)\citenamefont {Cho},
  \citenamefont {McCarron}, \citenamefont {Jenkin}, \citenamefont
  {K{\"{o}}ppinger},\ and\ \citenamefont {Cornish}}]{Cho2011}%
  \BibitemOpen
  \bibfield  {author} {\bibinfo {author} {\bibfnamefont {H.~W.}\ \bibnamefont
  {Cho}}, \bibinfo {author} {\bibfnamefont {D.~J.}\ \bibnamefont {McCarron}},
  \bibinfo {author} {\bibfnamefont {D.~L.}\ \bibnamefont {Jenkin}}, \bibinfo
  {author} {\bibfnamefont {M.~P.}\ \bibnamefont {K{\"{o}}ppinger}}, \ and\
  \bibinfo {author} {\bibfnamefont {S.~L.}\ \bibnamefont {Cornish}},\
  }\href@noop {} {\bibfield  {journal} {\bibinfo  {journal} {Eur. Phys. J. D}\
  }\textbf {\bibinfo {volume} {65}},\ \bibinfo {pages} {125} (\bibinfo {year}
  {2011})}\BibitemShut {NoStop}%
\bibitem [{\citenamefont {Lin}\ \emph {et~al.}(2009)\citenamefont {Lin},
  \citenamefont {Perry}, \citenamefont {Compton}, \citenamefont {Spielman},\
  and\ \citenamefont {Porto}}]{Lin:hybrid:2009}%
  \BibitemOpen
  \bibfield  {author} {\bibinfo {author} {\bibfnamefont {Y.-J.}\ \bibnamefont
  {Lin}}, \bibinfo {author} {\bibfnamefont {A.~R.}\ \bibnamefont {Perry}},
  \bibinfo {author} {\bibfnamefont {R.~L.}\ \bibnamefont {Compton}}, \bibinfo
  {author} {\bibfnamefont {I.~B.}\ \bibnamefont {Spielman}}, \ and\ \bibinfo
  {author} {\bibfnamefont {J.~V.}\ \bibnamefont {Porto}},\ }\href@noop {}
  {\bibfield  {journal} {\bibinfo  {journal} {Phys. Rev. A}\ }\textbf {\bibinfo
  {volume} {79}},\ \bibinfo {pages} {063631} (\bibinfo {year}
  {2009})}\BibitemShut {NoStop}%
\bibitem [{\citenamefont {Jenkin}\ \emph {et~al.}(2011)\citenamefont {Jenkin},
  \citenamefont {McCarron}, \citenamefont {K{\"{o}}ppinger}, \citenamefont
  {Cho}, \citenamefont {Hopkins},\ and\ \citenamefont {Cornish}}]{Jenkin:2011}%
  \BibitemOpen
  \bibfield  {author} {\bibinfo {author} {\bibfnamefont {D.~L.}\ \bibnamefont
  {Jenkin}}, \bibinfo {author} {\bibfnamefont {D.~J.}\ \bibnamefont
  {McCarron}}, \bibinfo {author} {\bibfnamefont {M.~P.}\ \bibnamefont
  {K{\"{o}}ppinger}}, \bibinfo {author} {\bibfnamefont {H.~W.}\ \bibnamefont
  {Cho}}, \bibinfo {author} {\bibfnamefont {S.~A.}\ \bibnamefont {Hopkins}}, \
  and\ \bibinfo {author} {\bibfnamefont {S.~L.}\ \bibnamefont {Cornish}},\
  }\href@noop {} {\bibfield  {journal} {\bibinfo  {journal} {Eur. Phys. J. D}\
  }\textbf {\bibinfo {volume} {65}},\ \bibinfo {pages} {11} (\bibinfo {year}
  {2011})}\BibitemShut {NoStop}%
\bibitem [{\citenamefont {Harris}\ \emph {et~al.}(2008)\citenamefont {Harris},
  \citenamefont {Tierney},\ and\ \citenamefont {Cornish}}]{Harris:2008}%
  \BibitemOpen
  \bibfield  {author} {\bibinfo {author} {\bibfnamefont {M.~L.}\ \bibnamefont
  {Harris}}, \bibinfo {author} {\bibfnamefont {P.}~\bibnamefont {Tierney}}, \
  and\ \bibinfo {author} {\bibfnamefont {S.~L.}\ \bibnamefont {Cornish}},\
  }\href@noop {} {\bibfield  {journal} {\bibinfo  {journal} {J. Phys. B -- At.
  Mol. Opt. Phys.}\ }\textbf {\bibinfo {volume} {41}},\ \bibinfo {pages}
  {035303} (\bibinfo {year} {2008})}\BibitemShut {NoStop}%
\bibitem [{\citenamefont {Wille}\ \emph {et~al.}(2008)\citenamefont {Wille},
  \citenamefont {Spiegelhalder}, \citenamefont {Kerner}, \citenamefont {Naik},
  \citenamefont {Trenkwalder}, \citenamefont {Hendl}, \citenamefont {Schreck},
  \citenamefont {Grimm}, \citenamefont {Tiecke}, \citenamefont {Walraven},
  \citenamefont {Kokkelmans}, \citenamefont {Tiesinga},\ and\ \citenamefont
  {Julienne}}]{Wille2008}%
  \BibitemOpen
  \bibfield  {author} {\bibinfo {author} {\bibfnamefont {E.}~\bibnamefont
  {Wille}}, \bibinfo {author} {\bibfnamefont {F.~M.}\ \bibnamefont
  {Spiegelhalder}}, \bibinfo {author} {\bibfnamefont {G.}~\bibnamefont
  {Kerner}}, \bibinfo {author} {\bibfnamefont {D.}~\bibnamefont {Naik}},
  \bibinfo {author} {\bibfnamefont {A.}~\bibnamefont {Trenkwalder}}, \bibinfo
  {author} {\bibfnamefont {G.}~\bibnamefont {Hendl}}, \bibinfo {author}
  {\bibfnamefont {F.}~\bibnamefont {Schreck}}, \bibinfo {author} {\bibfnamefont
  {R.}~\bibnamefont {Grimm}}, \bibinfo {author} {\bibfnamefont {T.~G.}\
  \bibnamefont {Tiecke}}, \bibinfo {author} {\bibfnamefont {J.~T.~M.}\
  \bibnamefont {Walraven}}, \bibinfo {author} {\bibfnamefont {S.~J. J. M.~F.}\
  \bibnamefont {Kokkelmans}}, \bibinfo {author} {\bibfnamefont
  {E.}~\bibnamefont {Tiesinga}}, \ and\ \bibinfo {author} {\bibfnamefont
  {P.~S.}\ \bibnamefont {Julienne}},\ }\href {\doibase
  10.1103/PhysRevLett.100.053201} {\bibfield  {journal} {\bibinfo  {journal}
  {Phys. Rev. Lett.}\ }\textbf {\bibinfo {volume} {100}},\ \bibinfo {pages}
  {053201} (\bibinfo {year} {2008})}\BibitemShut {NoStop}%
\bibitem [{\citenamefont {Moerdijk}\ \emph {et~al.}(1995)\citenamefont
  {Moerdijk}, \citenamefont {Verhaar},\ and\ \citenamefont
  {Axelsson}}]{Moerdijk:1995}%
  \BibitemOpen
  \bibfield  {author} {\bibinfo {author} {\bibfnamefont {A.~J.}\ \bibnamefont
  {Moerdijk}}, \bibinfo {author} {\bibfnamefont {B.~J.}\ \bibnamefont
  {Verhaar}}, \ and\ \bibinfo {author} {\bibfnamefont {A.}~\bibnamefont
  {Axelsson}},\ }\href@noop {} {\bibfield  {journal} {\bibinfo  {journal}
  {Phys. Rev. A}\ }\textbf {\bibinfo {volume} {51}},\ \bibinfo {pages} {4852}
  (\bibinfo {year} {1995})}\BibitemShut {NoStop}%
\bibitem [{\citenamefont {Fedichev}\ \emph {et~al.}(1996)\citenamefont
  {Fedichev}, \citenamefont {Reynolds},\ and\ \citenamefont
  {Shlyapnikov}}]{Fedichev:a4:1996}%
  \BibitemOpen
  \bibfield  {author} {\bibinfo {author} {\bibfnamefont {P.~O.}\ \bibnamefont
  {Fedichev}}, \bibinfo {author} {\bibfnamefont {M.~W.}\ \bibnamefont
  {Reynolds}}, \ and\ \bibinfo {author} {\bibfnamefont {G.~V.}\ \bibnamefont
  {Shlyapnikov}},\ }\href {\doibase 10.1103/PhysRevLett.77.2921} {\bibfield
  {journal} {\bibinfo  {journal} {Phys. Rev. Lett.}\ }\textbf {\bibinfo
  {volume} {77}},\ \bibinfo {pages} {2921} (\bibinfo {year}
  {1996})}\BibitemShut {NoStop}%
\bibitem [{\citenamefont {Wang}\ and\ \citenamefont {Julienne}()}]{Wang2013}%
  \BibitemOpen
  \bibfield  {author} {\bibinfo {author} {\bibfnamefont {Y.}~\bibnamefont
  {Wang}}\ and\ \bibinfo {author} {\bibfnamefont {P.~S.}\ \bibnamefont
  {Julienne}},\ }\href@noop {} {\bibinfo  {journal} {(private communication)}\
  }\BibitemShut {NoStop}%
\bibitem [{\citenamefont {Hodby}\ \emph {et~al.}(2005)\citenamefont {Hodby},
  \citenamefont {Thompson}, \citenamefont {Regal}, \citenamefont {Greiner},
  \citenamefont {Wilson}, \citenamefont {Jin}, \citenamefont {Cornell},\ and\
  \citenamefont {Wieman}}]{Hodby:2005}%
  \BibitemOpen
\bibfield  {journal} {  }\bibfield  {author} {\bibinfo {author} {\bibfnamefont
  {E.}~\bibnamefont {Hodby}}, \bibinfo {author} {\bibfnamefont {S.~T.}\
  \bibnamefont {Thompson}}, \bibinfo {author} {\bibfnamefont {C.~A.}\
  \bibnamefont {Regal}}, \bibinfo {author} {\bibfnamefont {M.}~\bibnamefont
  {Greiner}}, \bibinfo {author} {\bibfnamefont {A.~C.}\ \bibnamefont {Wilson}},
  \bibinfo {author} {\bibfnamefont {D.~S.}\ \bibnamefont {Jin}}, \bibinfo
  {author} {\bibfnamefont {E.~A.}\ \bibnamefont {Cornell}}, \ and\ \bibinfo
  {author} {\bibfnamefont {C.~E.}\ \bibnamefont {Wieman}},\ }\href@noop {}
  {\bibfield  {journal} {\bibinfo  {journal} {Phys. Rev. Lett.}\ }\textbf
  {\bibinfo {volume} {94}},\ \bibinfo {pages} {120402} (\bibinfo {year}
  {2005})}\BibitemShut {NoStop}%
\bibitem [{\citenamefont {Mark}\ \emph {et~al.}(2007)\citenamefont {Mark},
  \citenamefont {Ferlaino}, \citenamefont {Knoop}, \citenamefont {Danzl},
  \citenamefont {Kraemer}, \citenamefont {Chin}, \citenamefont {N\"agerl},\
  and\ \citenamefont {Grimm}}]{Mark:spect:2007}%
  \BibitemOpen
  \bibfield  {author} {\bibinfo {author} {\bibfnamefont {M.}~\bibnamefont
  {Mark}}, \bibinfo {author} {\bibfnamefont {F.}~\bibnamefont {Ferlaino}},
  \bibinfo {author} {\bibfnamefont {S.}~\bibnamefont {Knoop}}, \bibinfo
  {author} {\bibfnamefont {J.~G.}\ \bibnamefont {Danzl}}, \bibinfo {author}
  {\bibfnamefont {T.}~\bibnamefont {Kraemer}}, \bibinfo {author} {\bibfnamefont
  {C.}~\bibnamefont {Chin}}, \bibinfo {author} {\bibfnamefont {H.-C.}\
  \bibnamefont {N\"agerl}}, \ and\ \bibinfo {author} {\bibfnamefont
  {R.}~\bibnamefont {Grimm}},\ }\href {\doibase 10.1103/PhysRevA.76.042514}
  {\bibfield  {journal} {\bibinfo  {journal} {Phys. Rev. A}\ }\textbf {\bibinfo
  {volume} {76}},\ \bibinfo {pages} {042514} (\bibinfo {year}
  {2007})}\BibitemShut {NoStop}%
\bibitem [{\citenamefont {Herbig}\ \emph {et~al.}(2003)\citenamefont {Herbig},
  \citenamefont {Kraemer}, \citenamefont {Mark}, \citenamefont {Weber},
  \citenamefont {Chin}, \citenamefont {N\"{a}gerl},\ and\ \citenamefont
  {Grimm}}]{Herbig:2003}%
  \BibitemOpen
  \bibfield  {author} {\bibinfo {author} {\bibfnamefont {J.}~\bibnamefont
  {Herbig}}, \bibinfo {author} {\bibfnamefont {T.}~\bibnamefont {Kraemer}},
  \bibinfo {author} {\bibfnamefont {M.}~\bibnamefont {Mark}}, \bibinfo {author}
  {\bibfnamefont {T.}~\bibnamefont {Weber}}, \bibinfo {author} {\bibfnamefont
  {C.}~\bibnamefont {Chin}}, \bibinfo {author} {\bibfnamefont {H.~C.}\
  \bibnamefont {N\"{a}gerl}}, \ and\ \bibinfo {author} {\bibfnamefont
  {R.}~\bibnamefont {Grimm}},\ }\href@noop {} {\bibfield  {journal} {\bibinfo
  {journal} {Science}\ }\textbf {\bibinfo {volume} {301}},\ \bibinfo {pages}
  {1510} (\bibinfo {year} {2003})}\BibitemShut {NoStop}%
\bibitem [{\citenamefont {Mukaiyama}\ \emph {et~al.}(2004)\citenamefont
  {Mukaiyama}, \citenamefont {Abo-Shaeer}, \citenamefont {Xu}, \citenamefont
  {Chin},\ and\ \citenamefont {Ketterle}}]{Mukaiyama:2004}%
  \BibitemOpen
  \bibfield  {author} {\bibinfo {author} {\bibfnamefont {T.}~\bibnamefont
  {Mukaiyama}}, \bibinfo {author} {\bibfnamefont {J.~R.}\ \bibnamefont
  {Abo-Shaeer}}, \bibinfo {author} {\bibfnamefont {K.}~\bibnamefont {Xu}},
  \bibinfo {author} {\bibfnamefont {J.~K.}\ \bibnamefont {Chin}}, \ and\
  \bibinfo {author} {\bibfnamefont {W.}~\bibnamefont {Ketterle}},\ }\href@noop
  {} {\bibfield  {journal} {\bibinfo  {journal} {Phys. Rev. Lett.}\ }\textbf
  {\bibinfo {volume} {92}},\ \bibinfo {pages} {180402} (\bibinfo {year}
  {2004})}\BibitemShut {NoStop}%
\bibitem [{\citenamefont {D\"urr}\ \emph {et~al.}(2004)\citenamefont {D\"urr},
  \citenamefont {Volz},\ and\ \citenamefont {Rempe}}]{Durr:diss:2004}%
  \BibitemOpen
  \bibfield  {author} {\bibinfo {author} {\bibfnamefont {S.}~\bibnamefont
  {D\"urr}}, \bibinfo {author} {\bibfnamefont {T.}~\bibnamefont {Volz}}, \ and\
  \bibinfo {author} {\bibfnamefont {G.}~\bibnamefont {Rempe}},\ }\href@noop {}
  {\bibfield  {journal} {\bibinfo  {journal} {Phys. Rev. A}\ }\textbf {\bibinfo
  {volume} {70}},\ \bibinfo {pages} {031601} (\bibinfo {year}
  {2004})}\BibitemShut {NoStop}%
\bibitem [{\citenamefont {Danzl}\ \emph {et~al.}(2008)\citenamefont {Danzl},
  \citenamefont {Haller}, \citenamefont {Gustavsson}, \citenamefont {Mark},
  \citenamefont {Hart}, \citenamefont {Bouloufa}, \citenamefont {Dulieu},
  \citenamefont {Ritsch},\ and\ \citenamefont {N\"agerl}}]{Danzl:v73:2008}%
  \BibitemOpen
  \bibfield  {author} {\bibinfo {author} {\bibfnamefont {J.~G.}\ \bibnamefont
  {Danzl}}, \bibinfo {author} {\bibfnamefont {E.}~\bibnamefont {Haller}},
  \bibinfo {author} {\bibfnamefont {M.}~\bibnamefont {Gustavsson}}, \bibinfo
  {author} {\bibfnamefont {M.~J.}\ \bibnamefont {Mark}}, \bibinfo {author}
  {\bibfnamefont {R.}~\bibnamefont {Hart}}, \bibinfo {author} {\bibfnamefont
  {N.}~\bibnamefont {Bouloufa}}, \bibinfo {author} {\bibfnamefont
  {O.}~\bibnamefont {Dulieu}}, \bibinfo {author} {\bibfnamefont
  {H.}~\bibnamefont {Ritsch}}, \ and\ \bibinfo {author} {\bibfnamefont {H.-C.}\
  \bibnamefont {N\"agerl}},\ }\href@noop {} {\bibfield  {journal} {\bibinfo
  {journal} {Science}\ }\textbf {\bibinfo {volume} {321}},\ \bibinfo {pages}
  {1062} (\bibinfo {year} {2008})}\BibitemShut {NoStop}%
\bibitem [{\citenamefont {Debatin}\ \emph {et~al.}(2011)\citenamefont
  {Debatin}, \citenamefont {Takekoshi}, \citenamefont {Rameshan}, \citenamefont
  {Reichs\"ollner}, \citenamefont {Ferlaino}, \citenamefont {Grimm},
  \citenamefont {Vexiau}, \citenamefont {Bouloufa}, \citenamefont {Dulieu},\
  and\ \citenamefont {N\"agerl}}]{Debatin:2011}%
  \BibitemOpen
  \bibfield  {author} {\bibinfo {author} {\bibfnamefont {M.}~\bibnamefont
  {Debatin}}, \bibinfo {author} {\bibfnamefont {T.}~\bibnamefont {Takekoshi}},
  \bibinfo {author} {\bibfnamefont {R.}~\bibnamefont {Rameshan}}, \bibinfo
  {author} {\bibfnamefont {L.}~\bibnamefont {Reichs\"ollner}}, \bibinfo
  {author} {\bibfnamefont {F.}~\bibnamefont {Ferlaino}}, \bibinfo {author}
  {\bibfnamefont {R.}~\bibnamefont {Grimm}}, \bibinfo {author} {\bibfnamefont
  {R.}~\bibnamefont {Vexiau}}, \bibinfo {author} {\bibfnamefont
  {N.}~\bibnamefont {Bouloufa}}, \bibinfo {author} {\bibfnamefont
  {O.}~\bibnamefont {Dulieu}}, \ and\ \bibinfo {author} {\bibfnamefont {H.-C.}\
  \bibnamefont {N\"agerl}},\ }\href@noop {} {\bibfield  {journal} {\bibinfo
  {journal} {Phys. Chem. Chem. Phys.}\ }\textbf {\bibinfo {volume} {13}},\
  \bibinfo {pages} {18926} (\bibinfo {year} {2011})}\BibitemShut {NoStop}%
\bibitem [{\citenamefont {Aikawa}\ \emph {et~al.}(2011)\citenamefont {Aikawa},
  \citenamefont {Kobayashi}, \citenamefont {Oasa}, \citenamefont {Kishimoto},
  \citenamefont {Ueda},\ and\ \citenamefont {Inouye}}]{Aikawa2011}%
  \BibitemOpen
  \bibfield  {author} {\bibinfo {author} {\bibfnamefont {K.}~\bibnamefont
  {Aikawa}}, \bibinfo {author} {\bibfnamefont {J.}~\bibnamefont {Kobayashi}},
  \bibinfo {author} {\bibfnamefont {K.}~\bibnamefont {Oasa}}, \bibinfo {author}
  {\bibfnamefont {T.}~\bibnamefont {Kishimoto}}, \bibinfo {author}
  {\bibfnamefont {M.}~\bibnamefont {Ueda}}, \ and\ \bibinfo {author}
  {\bibfnamefont {S.}~\bibnamefont {Inouye}},\ }\href {\doibase
  10.1364/OE.19.014479} {\bibfield  {journal} {\bibinfo  {journal} {Opt.
  Express}\ }\textbf {\bibinfo {volume} {19}},\ \bibinfo {pages} {14479}
  (\bibinfo {year} {2011})}\BibitemShut {NoStop}%
\end{thebibliography}

%

\end{document}